\documentclass[usenatbib,longauth]{mn2e}

\usepackage{natbib}
\usepackage{longtable,lscape}
\usepackage{amsmath}
\bibpunct{(}{)}{;}{a}{}{,}
\usepackage{graphicx}

\usepackage{graphicx}
\usepackage[colorlinks=true,citecolor=blue]{hyperref}
\usepackage{color}
\usepackage{xspace}
\usepackage{dcolumn}
\usepackage{longtable}
\usepackage{lscape}
\usepackage{supertabular,booktabs}
\usepackage{url}
\usepackage{afterpage}

\usepackage{chemformula}

\newcommand{\MJup}{M$_{\mathrm{Jup}}$\xspace}
\newcommand{\RJup}{R$_{\mathrm{Jup}}$\xspace}

\newcommand{\MSun}{M$_{\odot}$\xspace}

\newcommand{\mic}{$\mu$m\xspace}
\newcommand{\as}{\hbox{$^{\prime\prime}$}\xspace}

\title[]{Characterising brown dwarf companions with IRDIS long-slit
  spectroscopy: HD\,1160\,B and HD\,19467\,B\thanks{Based on observations collected at the European Organisation for Astronomical Research in the Southern Hemisphere under ESO programme 102.C-0191(A).}}
\author[D. Mesa et al.]{D. Mesa$^{1}$\thanks{E-mail:
    dino.mesa@inaf.it (AVR)}, V. D'Orazi$^{1,2}$, A. Vigan$^{3}$,
  D. Kitzmann$^{4}$, K. Heng$^{4,9}$,
  R. Gratton$^{1}$, \newauthor S. Desidera$^{1}$, M. Bonnefoy$^{5}$, 
  B. Lavie$^{6}$, A.-L. Maire$^{7}$, S. Peretti$^{6}$, A. Boccaletti$^{8}$ \\ 
  $^{1}$INAF-Osservatorio Astronomico di Padova, Vicolo dell'Osservatorio 5, Padova, Italy, 35122-I\\
  $^{2}$ School of Physics and Astronomy, Monash University, Clayton campus, Victoria 3800, Melbourne, Australia\\ 
  $^{3}$Aix Marseille Univ., CNRS, CNES, LAM, Marseille, France \\
  $^{4}$Center for Space and Habitability, University of Bern, Gesellschaftsstrasse 6, Bern, 3012, Switzerland \\
  $^{5}$Univ. Grenoble Alpes, CNRS, IPAG, 38000 Grenoble, France\\
  $^{6}$Geneva Observatory, University of Geneva, Chemin des Maillettes 51, 1290 Versoix, Switzerland \\
  $^{7}$STAR Institute, Universit\'{e} de Li\`ege, All\'{e}e du Six Ao\^{u}t 19c, B-4000, Li\`ege, Belgium \\
  $^{8}$LESIA, Observatoire de Paris, Universit\'{e} PSL, CNRS, Sorbonne
  Universit\'{e}, Univ. Paris Diderot, Sorbonne Paris Cit\'{e}, 5 place Jules
  \\ Janssen, F-92195 Meudon, France \\
  $^{9}$University of Warwick, Department of Physics, Astronomy \& Astrophysics Group, Coventry CV4 7AL, U.K.\\
}
\begin{document}

\date{Accepted . Received ; in original form }

\pagerange{\pageref{firstpage}--\pageref{lastpage}} \pubyear{}

\maketitle

\label{firstpage}

\begin{abstract}
  The determination of the fundamental properties 
  (mass, separation, age, gravity and atmospheric properties) of brown dwarf
  companions allows us to infer crucial informations on their formation and
  evolution mechanisms. Spectroscopy of substellar companions is available to
  date only for a limited number of objects (and mostly at very low resolution,
  R$<$50) because of technical limitations, i.e., contrast and angular
  resolution. We present medium resolution (R=350), coronagraphic long-slit
  spectroscopic observations with SPHERE of two substellar companions,
  HD\,1160\,B and HD\,19467\,B. We found that HD\,1160\,B has a peculiar
  spectrum that cannot be fitted by spectra in current spectral libraries. A
  good fit is possible only considering separately the Y+J and the H spectral
  band. The spectral type is between M5 and M7. We also estimated
  a $T_{eff}$ of
  2800-2900~K and a $\log{g}$ of 3.5-4.0~{\it dex}. The low surface gravity
  seems to favour young age (10-20~Myr) and low mass ($\sim$20~\MJup) for this
  object. HD\,19467\,B is instead a fully evolved object with a $T_{eff}$ of
  $\sim$1000~K and $\log{g}$ of $\sim$5.0~{\it dex}. Its spectral type is 
  T6$\pm$1.
\end{abstract}


\section{Introduction}
\label{s:intro}

Brown dwarfs (BDs) are low-mass objects populating the mass range between stars and planets. Firstly theorised by \citet{1963ApJ...137.1126K} and \citet{1963PThPh..30..460H},  
it was not until 1995 that the first unquestionable evidence of their existence was provided \citep{rebolo95,1995Natur.378..463N}. Because of their low mass, these objects are not capable to sustain hydrogen burning in their cores and they continuously cool down during their lifetime \citep{baraffe2003}. The distinction between BDs and planets is conventionally set at the minimum mass for which deuterium (D) can fuse into $^3$He nuclei. This
value has been defined as the mass of objects able to burn the 90\% of their initial D in 10~Gyr and it was found to range from 13.1 to 12.4~\MJup, for cloudless and cloudy models, respectively \citep{2008ApJ...689.1327S}.
Furthermore, \citet{2011ApJ...727...57S} found that this value is valid for objects at solar metallicity, while it varies from 11.0~\MJup for a metallicity three times the solar value to 16.3~\MJup for zero metallicity. \par
An alternative approach to distinguish between planets and BDs is through their formation mechanism. The detection of sub-stellar objects in young stellar clusters \citep[e.g. ][]{2010ApJ...718..105D} and the fact that collapse
formation scenarios foresee that the minimum fragment masses are well extended in the sub-stellar regime \citep[e.g. ][]{2019MNRAS.490.3061V} are clear hints that the formation through the fragmentation of molecular clouds is
possible also for this type of objects. Alternatively, they could form following planet-like mechanisms like the gravitational instability
\citep{1978M&P....18....5C} into protoplanetary disks. It is nowadays
recognized that these two formation mechanisms overlap on mass ranges
of few Jupiter masses \citep{2019A&A...631A.125K}. \par
To clarify the nature of these objects,
studies both in the visible \citep{2009AJ....137.3345C} and in the near-infrared
\citep[NIR, e.g. ][]{2010ApJ...715..561A,2011AN....332..821S,
  2013ApJ...772...79A,2017ApJ...838...73M} were performed on field BDs.
Observing these objects in the NIR is very important because a large fraction
of the flux emitted by L and T dwarfs is at wavelenghts ranging between
1 and 2.5~\mic. In addition, the NIR spectrum of these objects contains a wealth
of spectral features ranging from narrow atomic lines to broad molecular bands
that have different dependencies on their temperature, gravity and metallicity
\citep{2005ARA&A..43..195K}. \par
In recent years, new high-contrast imaging
instruments like NACO \citep{2003SPIE.4841..944L}, GPI
\citep{2014AAS...22322902M}, NICI \citep{2008SPIE.7015E..1VC}, NIRI
\citep{2003PASP..115.1388H}, NIRC2 \citep[e..g. ][]{2017AJ....153...44M},
and SPHERE \citep{2019A&A...631A.155B} allowed
to detect several BDs companions of main sequence or pre-main sequence stars
\citep[see e.g., ][]{2015ApJ...811..103M,2016ApJ...829L...4K,2017A&A...597L...2M,2018A&A...615A.160C}. The possibility to study the spectral characteristics of these BD companions can provide additional, complementary information to observations of isolated, field BDs.
\par
Currently, only a handful of BD companions has been spectroscopically investigated exploiting the limited
resolution (R$\sim$30-50) of integral-field spectrograph such as e.g., SPHERE and GPI. Some examples
include HR\,2562\,B \citep{2016ApJ...829L...4K,2018A&A...612A..92M},
HD\,984\,B \citep{2017AJ....153..190J}, HD\,1160\,B
\citep{2016A&A...587A..56M,2017ApJ...834..162G}, HD\,206893\,B
\citep{2017A&A...608A..79D}, HD\,4747\,B
\citep{2018ApJ...853..192C,2019A&A...631A.107P} and
HD\,13724\,B \citep{2020arXiv200208319R}. In only five more cases it was
possible to obtain a spectrum with a larger resolution (R=350), using the
SPHERE long-slit spectroscopy \cite[LSS -][]{2008A&A...489.1345V} observing
mode. This has been done for 2MASS\,J01225093-2439505\,B
\citep{2015ApJ...805L..10H}, PZ\,Tel\,B \citep{2016A&A...587A..56M},
HR\,3549\,B \citep{2016A&A...593A.119M}, HD\,284149\,B
\citep{2017A&A...608A.106B} and HIP\,64892\,B \citep{2018A&A...615A.160C}.
Moreover, spectra in the NIR for companions at large separation from the host
star were obtained at medium resolution (1500-2000) with SINFONI
\citep[e.g. ][]{2014A&A...562A.127B}. \par
In this paper we expand upon the current sample and we focus on two sub-stellar objects orbiting the nearby stars HD\,1160 and HD\,19467. These systems are described in detail in Section~\ref{s:target}. Furthermore, we present in
Section~\ref{s:data} our data and the data reduction procedure adopted.
In Section~\ref{s:result}, we then present and discuss our main
results. Finally, in Section~\ref{s:conclusion} we give our conclusions.

\section[]{Target properties}
\label{s:target}

HD\,1160 is an A0V \citep{1999MSS...C05....0H} star at a distance of
125.9$\pm$1.2~pc \citep{2018yCat.1345....0G}. Around this star,
\citet{2012ApJ...750...53N} detected  two companions, HD\,1160\,B and
HD\,1160\,C at separations of $\sim$80~au and 530~au, respectively, 
using NICI at the Gemini telescope. They found a spectral type of L0
for HD\,1160\,B and a spectral type of M3.5 for HD\,1160\,C. They were also
able to determine a mass in the range between 24 and 45~\MJup, in the BDs
regime, for the former and in the range between  0.18 and 0.25~\MSun, in the
stellar regime, for the latter assuming an age between 10 and 100~Myr. \par
HD\,1160 was also observed with SPHERE in IRDIFS\_EXT mode obtaining a low
resolution (R=30) spectrum of HD\,1160\,B in the NIR between 0.95 and
1.65~\mic \citep{2016A&A...587A..56M}. This allowed to re-classify its spectral
type as M6 and to derive a $T_{eff}$ of 3000$\pm$100~K and a subsolar
metallicity.  However, they could not provide constraints on the surface gravity. They assumed a wider range for the age (30-300~Myr) resulting in masses between 39 and 166~\MJup. \par
More recently, this system was studied trough SCExAO and GPI by \citet{2017ApJ...834..162G}. They found for
HD\,1160\,B $T_{eff}$=3000-3100~K and $\log{g}$=4-4.5~{\it dex}, in agreement with the relatively young age of the system. Differently from \citet{2016A&A...587A..56M}, they did not find evidence of subsolar metallicity, as expected according to standard Galactic chemical evolution models and observations (see e.g., \citealt{bensby2014}; \citealt{spina2017}, and references therein). The authors also clearly stressed that the main uncertainty on the mass determination for this object is due to the estimation of the age of the system (for ages between 80 and 125~Myr, the mass of this object can vary from the BD regime to hydrogen-burning limit). \par
Recently, \citet{2019AJ....158...77C} assigned HD\,1160 as a member of the
recently discovered Pisces-Eridanus (Psc-Eri) stellar stream
\citep{2019A&A...622L..13M} exploiting kinematics. They also
defined for the stream an age of 120~Myr, in fair agreement with the estimate of  135~Myr later found by \citet{2020arXiv200203610R}. Should the membership of
HD\,1160 to the Psc-Eri stream be confirmed, the mass of HD\,1160\,B would be of the order of 0.12~\MSun, well into the stellar mass range \citep{2019AJ....158...77C}. \par
HD\,19467 is a G3 \citep{1988mcts.book.....H} star at a distance of
32.02$\pm$0.04~pc \citep{2018yCat.1345....0G}. Using NIRC2 high-contrast imaging
and HIRES radial velocity (RV) data, \citet{2014ApJ...781...29C} detected a
companion at a separation larger than 1.6\as. Combining Doppler observations and
imaging, they estimated a minimum mass just above 50~\MJup for the companion.
Moreover, they estimated an age of 4.3~Gyr through gyrochronology while isochrones provided an older age of 9$\pm$1~Gyr. By adopting these ages and using $K_s$
photometry, \citet{2014ApJ...781...29C} obtained companion masses
of 56.7~\MJup and 67.4~\MJup, respectively. Finally, from the measured colors they inferred a T5-T7 spectral type. \par
HD\,19467 was also
observed using the Project~1640 instrument at the Palomar Observatory by
\citet{2015ApJ...798L..43C}. They extracted a low resolution (R$\sim$30)
spectrum of HD\,19467\,B that allowed to define a spectral type of T5.5$\pm$1
and a $T_{eff}$ of 980~K. \par
Recently, \citet{2019ApJ...873...83W} estimated,
through isochronal models, an age of $10.06^{+1.16}_{-0.82}$~Gyr for HD\,19467, and found a mass of 67-69~\MJup for HD\,19467\,B. 
Very recently, Maire et al. (submitted) presented a thorough investigation of the host star properties and found for the system an age of $8.0_{-1.0}^{+2.0}$~Gyr.
We will then assume in our work this age estimation for consistency with that
paper.
  
\section[]{Observations and data reduction}
\label{s:data}

We observed  HD\,1160 and HD\,19467 with the SPHERE IRDIS
\citep{2008SPIE.7014E..3LD} scientific subsystem operating in the LSS medium resolution (MRS) observing mode. This mode allows to obtain spectra
with a resolution of R=350 covering the Y, J and H spectral bands simultaneously
for companions with contrasts of the order of $10^{-5}-10^{-6}$ at separations
larger than 0.5\as. \par
We observed HD\,1160 in the night of 2018-08-20 and HD\,19467 in the
night of 2019-01-26. Both these observations were obtained in service
mode. In Table~\ref{t:obs} we report the main informations regarding the two
observing nights. In both nights seeing and coherence time were stable
along all the observations resulting in a stable Strehl ratio with values above
80\%. In the same Table we also listed the number of datacubes obtained for
  each observation together with the number of frames obtained for each
  datacube and the exposure time for each frame.
The total observing time on the target was of 24 minutes for
HD\,1160 and around 30 minutes for HD\,19467. In both cases we alternated observations with the slit on the position of the companion and observations with the slit rotated outside the position of the companion. This observing techniques
was first proposed for the case of the SPHERE LSS observing mode in
\citet{2016SPIE.9912E..26V} and its aim is to use the
data with the slit outside the companion to create a
reliable model of the speckle pattern to be subtracted from the data with
the slit on the companion. All the data
(i.e., with/without the companion into the slit)
were taken with the host star behind the coronagraph. In order to properly calibrate the flux of the extracted spectrum of the companion, we also obtained a spectrum of the star outside the coronagraphic mask. Finally we observed, in the same configuration
of the flux calibration data, one early-type star for each of our target just after the end of the scientific observations. They were then used to remove telluric lines and obtain a more accurate
wavelength solution. To this aim we observed the B8 star HD\,225187 in the case of HD\,1160 and the B3 star HD\,20001 in the case of HD\,19467. \par
We prepared a custom Python-based pipeline to reduce these data (detailed information will be provided in a forthcoming publication Mesa et al., in preparation).  As first
step, master dark, bad pixels map and master detector
flat are produced. We then create the wavelength calibration file. The scientific data with
the companion in the slit and those with the companion outside the slit are
then reduced separately applying the appropriate calibrations and finely
registering each frame. We also corrected a known problem of the MRS observing
mode that, due to a slight tilt ($\sim$1 degree) of the grism on its mount,
produce a different position of the companion point spread function (PSF) at
different wavelengths. The reduced data without the companion were then used
to create a model of the speckle pattern using a procedure based on the
principal components analysis \citep[PCA; ][]{2012ApJ...755L..28S}.
This model was then subtracted from each frame of the dataset with the
companion into the slit to effectively subtract the speckle pattern from them
and to improve the signal to noise ratio (SNR) of the extracted spectrum that
is composed by 780 wavelengths ranging between 0.94 and 1.82~\mic. For the
spectral extraction we used a fixed extraction window following the method
devised in \citet{2017A&A...608A.106B}. The uncertainties on the extracted
spectrum were obtained calculating the standard deviation on a window defined
in the same way of that used for the spectrum extraction. Moreover, this window
was at the same separation of the companion but on the opposite side with
respect to the host star. We extracted also the spectra of the
host star and of the standard star following the same method used for the
companion. We then finely recalibrated the wavelengths of the companion
and of the host star spectra using as reference the positions of the telluric
lines found in the standard star. Finally we obtained the spectrum of the
companion in contrast dividing it by the spectrum of the host star. \par
With the aim to check the reliability of our results we also reduce
our data using the SILSS pipeline \citep{2016ascl.soft03001V} which is
purposely designed to reduce SPHERE LSS datasets. 
Both for HD\,1160 and for HD\,19467 the extracted spectra with the two
pipelines were in good agreement confirming the reliability of our results.
To quantify the agreement between the two methods, we calculated the
differences between the two spectra and we verified that, for each
wavelength, this is smaller than the uncertainties.

\begin{table*}
 \centering
 \begin{minipage}{140mm}
   \caption{Characteristics of the SPHERE observations presented in this work.
     In column 1 we list the observing night, in column 2 we list the target
     name and in column 3 we report the coronagraph used. In columns
     4 and 5 we list the number of datacubes, the number of frames for each
     datacube and the exposure time expressed in s for each frame for
     observation with the slit on (labeled Obs. comp.) and outside the
     companion position (labeled Obs. no comp.). In columns 6, 7 and 8 we
     report the median values of seeing, coherence time and wind speed during
     the observations. \label{t:obs}}
  \begin{tabular}{c c c c c c c c}
  \hline
    Date & Target & Coronagraph &  Obs. comp.  & Obs. no comp. & S (\as) & $\tau_{0}$ (ms) & wind (m/s)\\
    \hline
         2018-08-20  & HD\,1160 & N\_S\_MR\_WL  &  5$\times$9;32  &  5$\times$8;32 &  0.49  & 4.6   &  11.10 \\
         2019-01-26  & HD\,19467 & N\_S\_MR\_WL  &  5$\times$11;32  &  5$\times$10;32 &  0.47  & 17.4  &  2.13\\
\hline
\end{tabular}
\end{minipage}
\end{table*}

\section{Results and discussion}
\label{s:result}

\subsection{HD\,1160\,B}
\label{s:HD1160res}

The contrast spectrum of the companion extracted following the method
described in Section~\ref{s:data} is transformed in flux multiplying it for
an appropriate BT-NEXTGEN \citep{2012RSPTA.370.2765A} theoretical spectrum,
following a standard procedure previously used in other works of our group
\citep[e.g. ][]{2016A&A...587A..55V,2016A&A...587A..57Z,2016A&A...587A..56M,2017A&A...608A.106B,2018A&A...612A..92M}. To identify the
appropriate model spectrum, we used the VOSA SED analyzer\footnote{\url{http://svo2.cab.inta-csic.es/theory/vosa/}} \citep{2008A&A...492..277B} to
find the model with the best fit with the stellar spectral energy distribution
(SED). At the end of this
procedure we selected a model with $T_{eff}$=9200~K, surface gravity
$\log{g}$=4.5~{\it dex} and solar metallicity. \par
\begin{figure}
\centering
\includegraphics[width=\columnwidth]{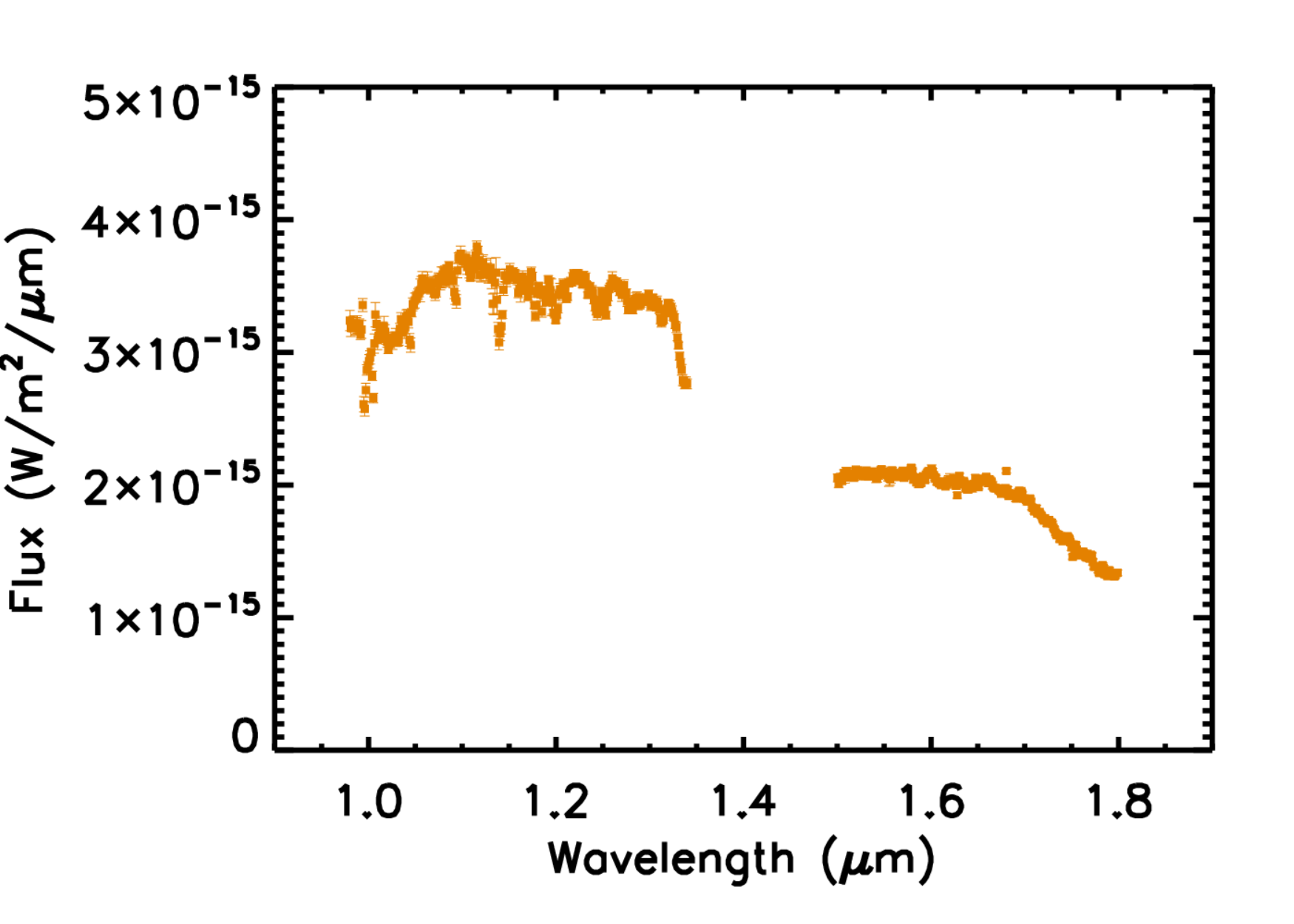}
\caption{Spectrum for HD\,1160\,B extracted using the method described in
  Section~\ref{s:HD1160res}.}
\label{f:HD1160spec}
\end{figure}

The spectrum resulting from this
procedure was very noisy for the lower and for the higher wavelengths and for
this reason we excluded all the wavelengths below 0.97~\mic and above 1.80~\mic.
Furthermore, we excluded also all the wavelengths in the spectral region
around 1.4~\mic that are affected by the presence of water telluric absorption.
At the end of this procedure, we remained with a
spectrum composed of 658 data points that is shown in Figure~\ref{f:HD1160spec}.
\par
From the extracted spectrum we obtained the photometric values for Y
(wavelengths below 1.15~\mic), J (wavelengths between 1.15 and 1.4~\mic) and H
(wavelengths above 1.4~\mic) spectral bands that are listed in the second
column of Table~\ref{t:allphoto}. Thanks to these values and exploiting the
BT-Settl models \citep{2014IAUS..299..271A}, we estimated the mass of the
companion. To this aim we adopted both a young system age of
$50_{-40}^{+50}$~Myr as proposed by \citet{2012ApJ...750...53N} and the older age of $\sim$120~Myr proposed by \citet{2019AJ....158...77C}. In this second case,
as an error on the age was not given by
\citet{2019AJ....158...77C}, we considered an uncertainty of 15~Myr using also the age determination from 
\citet{2020arXiv200203610R} for the Psc-Eri stellar stream. The results of this
procedure are listed in the second and the third column of
Table~\ref{t:allmass}. In the first case (i.e., young age), the uncertainties on the mass are very large mainly due to the large uncertainties on the system age, so that masses encompass the low-mass brown dwarfs/stellar regime.
As for the old age, the companion would be in the low-mass star regime, although at lower masses than those estimated by \citet{2019AJ....158...77C}. 

\begin{table}
 \centering
 \begin{minipage}{90mm}
   \caption{Absolute magnitude in Y, J and H band for HD\,1160 (second
     column) and HD\,19467 (third column). \label{t:allphoto}}
  \begin{tabular}{c c c}
  \hline
  Sp. Band &  Abs. Mag. (HD\,1160)  &  Abs. Mag. (HD\,19467) \\
         \hline
   Y &     9.92$\pm$0.01       &  16.39$\pm$0.07     \\
   J &     9.37$\pm$0.01       &  15.13$\pm$0.07     \\
   H &     9.01$\pm$0.01       &  15.84$\pm$0.08     \\
\hline
\end{tabular}
\end{minipage}
\end{table}

\begin{table*}
 \centering
 \begin{minipage}{140mm}
   \caption{Mass estimates in Y, J and H band for HD\,1160\,B considering the
     younger ages proposed by \citet{2012ApJ...750...53N} (second
     column - labeled as YA in the Table), for HD\,1160\,B considering
     the older age proposed by \citet{2019AJ....158...77C} (third column -
       labeled as OA in the Table) and HD\,19467
     (fourth column). \label{t:allmass}}
  \begin{tabular}{c c c c}
  \hline
  Sp. Band &  Mass (HD\,1160\,B - YA) \MJup  &  Mass (HD\,1160\,B - OA) \MJup & Mass (HD\,19467\,B) \MJup \\
         \hline
   Y &  $70.3_{-47.2}^{+27.5}$   &  $106.0_{-6.4}^{+6.1}$   &  $64.1_{-2.4}^{+2.0}$   \\
   J &  $70.4_{-49.4}^{+28.0}$   &  $107.5_{-7.1}^{+6.7}$   &  $66.4_{-1.9}^{+2.1}$   \\
   H &  $62.8_{-42.6}^{+25.2}$   &  $96.2_{-6.3}^{+4.8}$    &  $62.2_{-3.5}^{+2.5}$   \\
\hline
\end{tabular}
\end{minipage}
\end{table*}

The spectral type of HD\,1160\,B is obtained by fitting its extracted spectrum
with a sample of template spectra of field BDs taken from the
{\it Montreal Brown Dwarf and Exoplanet Spectral Library}\footnote{\url{https://jgagneastro.com/the-montreal-spectral-library/}}. To construct this
sample we have considered only objects for which spectra with a resolution
higher than that of the LSS mode of SPHERE were available. The spectra
were then convolved with a gaussian PSF to reduce the resolution to the
same value of the LSS spectra. Finally, we interpolated the flux values to the
same wavelengths of the LSS spectrum. \par
\begin{figure}
\centering
\includegraphics[width=\columnwidth]{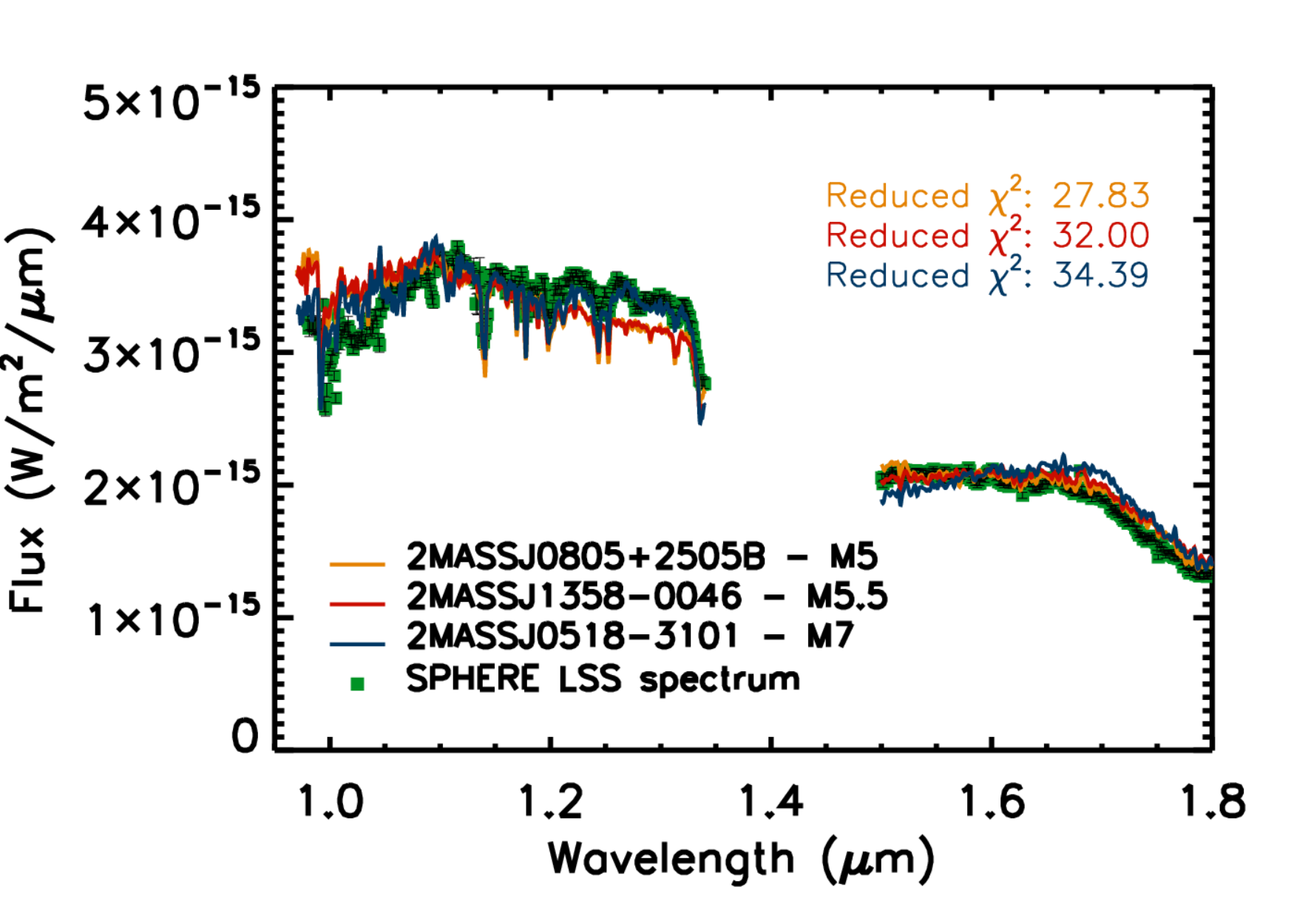}
\caption{Comparison of the extracted spectrum for HD\,1160\,B (green squares)
  with the three best fit spectra from the Montreal library (red, orange and
  blue solid lines).}
\label{f:HD1160spt}
\end{figure}

The best fit from this procedure is with the spectrum of 2MASSJ0805+2505\,B
a M5 spectral type field dwarf \citep{2015ApJS..219...33G}. Fits of
comparable quality were also obtained for template spectra with spectral types
M6 and M7 as shown in Figure~\ref{f:HD1160spt} where we also display the
spectrum of 2MASSJ1358+0046 a M5.5 spectral type \citep{2015ApJS..219...33G}
and of the M7 spectral type object 2MASSJ0518-3101 \citep{2015ApJS..219...33G}.
We can then conclude that the spectral type for HD\,1160\,B is M6$\pm$1. \par
However, it is important to note that the quality of the fits, as highlighted
by the value of the reduced $\chi^2$ is not good for any of the template
spectra. Indeed, none of them was able to produce a good fit both for the
Y+J and the H spectral bands. This result is not completely surprising as
similar difficulties to find a good fit with template spectra was experienced
also by \citet{2016A&A...587A..56M} as can be seen from Figure~12 in
that paper. Moreover, if we consider separately a fit
for the two spectral bands, we find that while the first one tends to
favour later (M7$\pm$1) spectral types, the latter favours later spectral
types (M5$\pm$1). In the second case the best fit is still with the M5 object
2MASSJ0805+2505\,B while in the first case the best fit is obtained with the
M7 spectral type 2MASSJ0518-3101. \par
\begin{figure}
\centering
\includegraphics[width=\columnwidth]{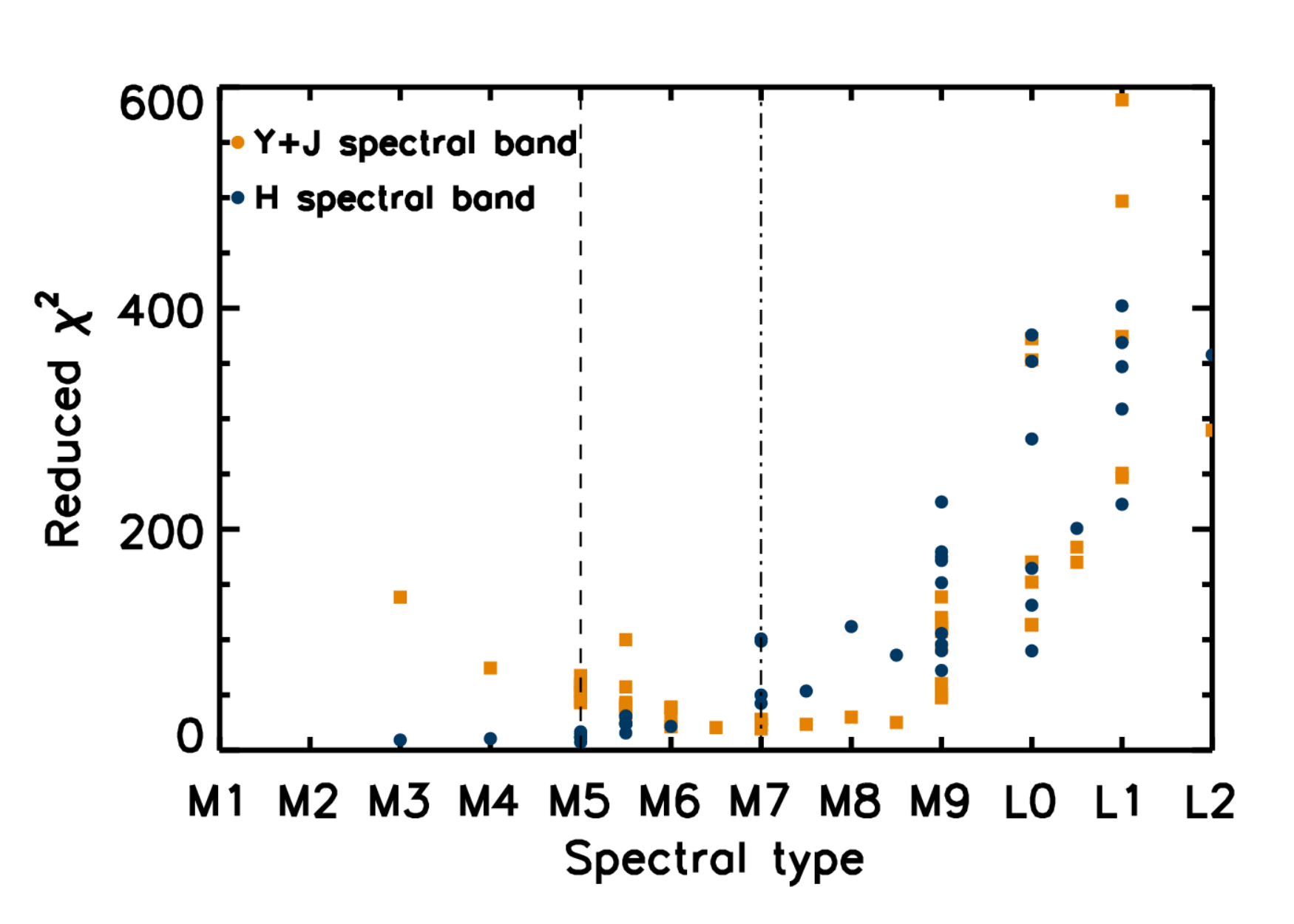}
\caption{Values of the $\chi^2$ obtained from the fit versus the spectral type
  of the template spectra. The orange squares represent the values obtained
  using only the Y+J spectral bands, the blue circles represent the values
  obtained using the H spectral band. The dashed vertical line highlights the
  position of the minimum for the H-band distribution, the dot-dashed line is
  the same thing for the Y+J-bands distribution.}
\label{f:HD1160chi}
\end{figure}

To further confirm this result
we also display in Figure~\ref{f:HD1160chi} the distributions of the $\chi^2$
values with respect to the spectral type of the fitting template spectra both
for the Y+J-bands and the H-band. As highlighted by the two vertical lines, it is apparent that the two distributions have two different minimum at
$\sim$M7 and $\sim$M5, respectively. While this difference is not large, it
is however an indication that HD\,1160\,B has a peculiar spectrum. \par
An alternative way to classify the spectral type of this substellar object is
to use spectral indices. To this aim we have used the indices described in
\citet{2013ApJ...772...79A} that are fitted to the wavelength band of our
observations. It is important to stress that these indices were thought for
spectra at resolution slightly higher ($\sim$700) than that used in our
analysis. While these results should be then taken with care, they can
however give a good indication of the spectral type of HD\,1160\,B.
In particular we used the $H_2O$ index \citep{2007ApJ...657..511A}
that is effective for spectral types between M5 and L4 with an rms of 0.390
in spectral type  and the $H_2O-1$ index \citep{2004ApJ...610.1045S} that is
effective for spectral types between M4 and L5 but with a larger rms of 1.097.
The $H_2O$ index, that is based on H-band wavelengths gives a spectral index
of M4.8 while the $H_2O$ index, that based on J-band wavelengths gives a later
spectral type of M6. \par
These results confirm what we found with the fit with
spectral templates and are a further indication of the peculiarity of
the spectrum of HD\,1160\,B. The reasons of this peculiarity are not
completely clear but we can do the hypothesis that it could be due to the
presence of dust in its photosphere or to its young age (few tens of Myr).
  The second possibility would imply a still not fully evolved object
  explaining its peculiar spectrum. The presence of dust
is in conflict with the high $T_{eff}$ found for this object while the young
age is in disagreement with the age of $\sim$120~Myr for the HD\,1160 system
proposed by \citet{2019AJ....158...77C}. \par
\begin{figure}
\centering
\includegraphics[width=\columnwidth]{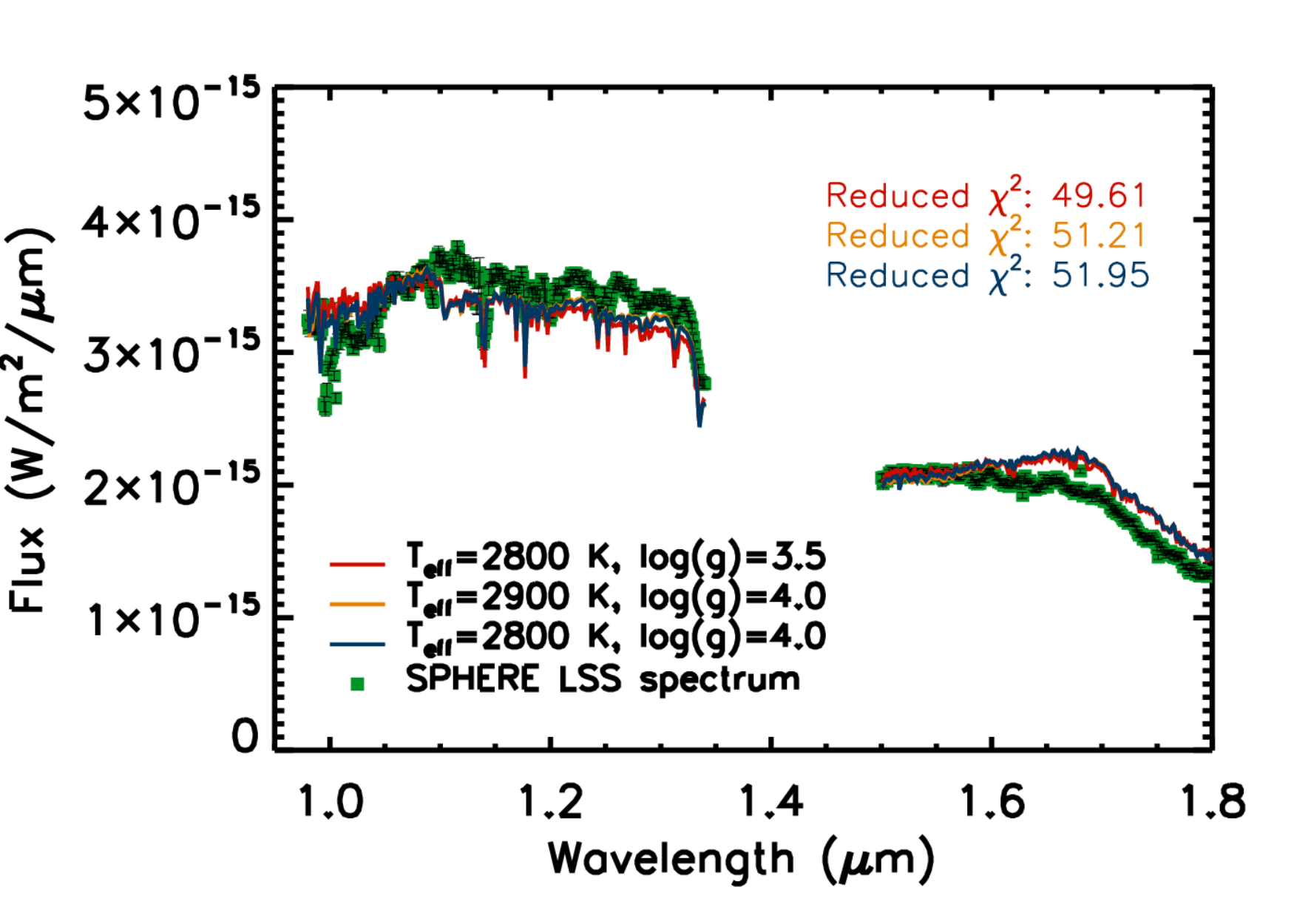}
\caption{Comparison of the extracted spectrum for HD\,1160\,B (green squares)
  with the three best fit theoretical BT-Settl spectra (orange, red and
  blue solid lines).}
\label{f:HD1160synt}
\end{figure}

With the aim to better define the physical characteristics of HD\,1160\,B, we
also fitted the extracted spectrum to a grid of BT-Settl atmospheric
models \citep{2014IAUS..299..271A} with a $T_{eff}$ ranging between 900 and
4000~K with a step of 100~K, with $\log{g}$ ranging between 0.0 and 6.0~{\it
  dex} with a step of 0.5~{\it dex} and a solar metallicity following what
previously found for this object by \citet{2017ApJ...834..162G}. The
results of this procedure are displayed in Figure~\ref{f:HD1160synt} where
the extracted spectrum of HD\,1160\,B is compared with the three best fit
models. The best fit is obtained for a model with a $T_{eff}$ of 2800~K and a
surface gravity of $\log{g}$=3.5~{\it dex} but comparably good fits are obtained
for $T_{eff}$ of 2900~K and a surface gravity of 3.5-4.0~{\it dex}. From this
analysis we conclude that HD\,1160\,B has $T_{eff}$=2800-2900~K and
$\log{g}$=3.5-4.0~{\it dex}. This $T_{eff}$ is in good agreement with the
spectral classification of M7 found through the comparison with the Y+J-bands
spectral templates \citep{2016MNRAS.461..794P}. 
\par
Like for the case of the fit with the template spectra, we note that
  it is not possible to obtain a completely satisfactory fit between the
  extracted spectrum and the atmospheric models. This is a further indication
  that the spectrum of HD\,1160\,B is peculiar and that its classification
  is difficult. Also in this case, possible explanations are the presence of
  dust and/or the youth of the system. \par
\begin{figure}
\centering
\includegraphics[width=\columnwidth]{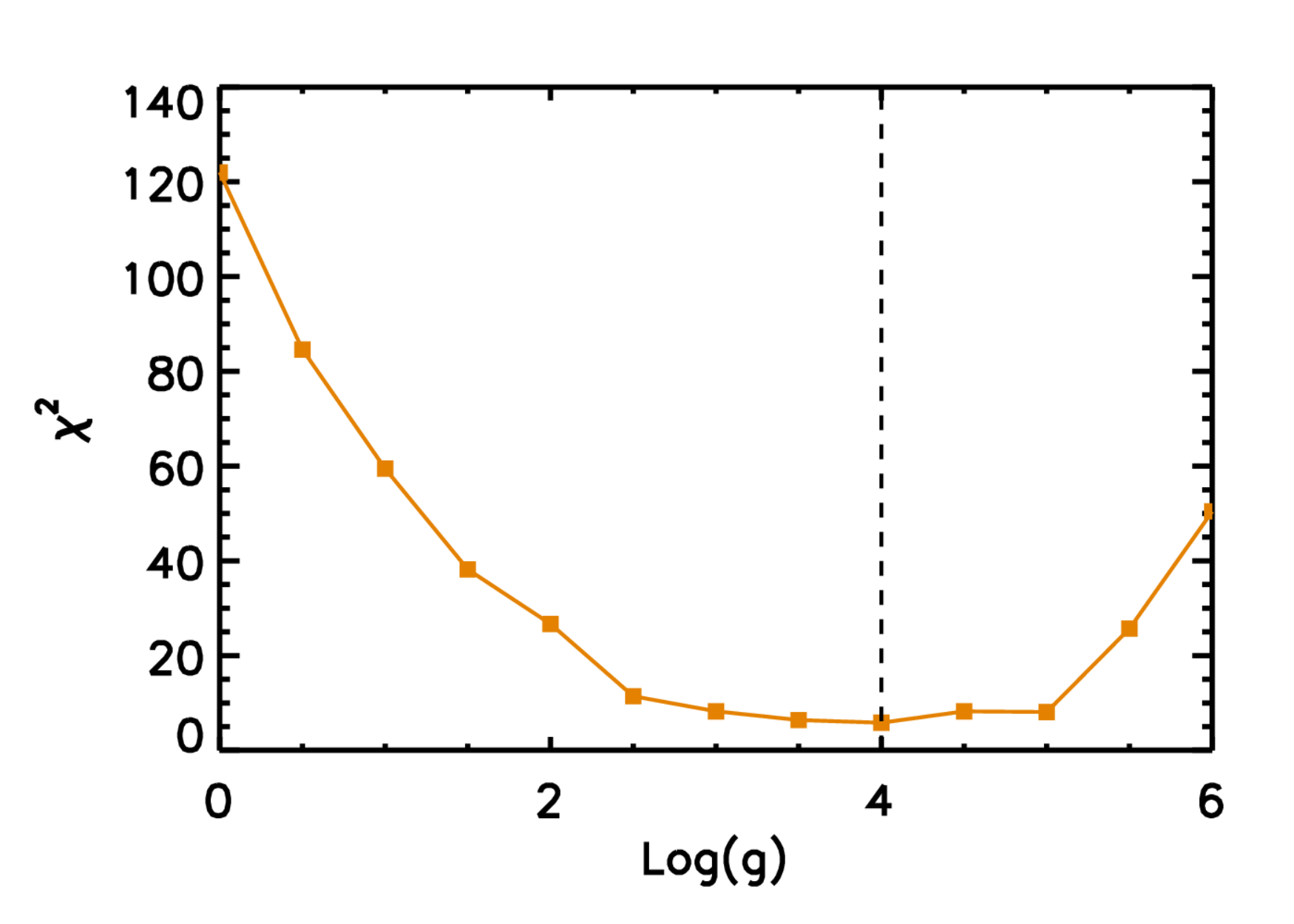}
\caption{Values of the $\chi^2$ obtained from the fit versus the values of
  $\log{g}$ for a fixed $T_{eff}$ of 2800~K. The vertical dashed line highlights
  the position of the minimum of the curve.}
\label{f:chigrav}
\end{figure}

  
To further constrain the gravity of HD\,1160\,B we considered the alkali lines
(Na~{\sc i} at 1.14~\mic and K~{\sc i} doublets at $\sim$1.17 and $\sim$1.25~\mic) that,
for a given $T_{eff}$, tend to become weaker at lower gravity
\citep[e.g. ][]{2003ApJ...593.1074G,2005ApJ...623.1115C}. To this aim we have
fitted the J-band (between 1.10 and 1.30~\mic) spectrum of HD\,1160\,B with
the atmospheric models with a fixed $T_{eff}$ of 2800~K. In
  Figure~\ref{f:chigrav} we show how the value of $\chi^2$ varies with
  $\log{g}$. The best fit is obtained for $\log{g}$ of 3.5-4.0~{\it dex}
confirming what we found with the fit on the whole spectrum and with different
$T_{eff}$. \par
Furthermore, we tested the gravity sensitive spectral index linked to the
Na~{\sc i} line at 1.14~\mic defined in \citet{2007ApJ...657..511A} for spectra
with a resolution of the order of 300 similar to that of our data. We obtained
for this index a value of 1.08 that indicates low-gravity. This index is
however reliable only for spectral types later than M6 so that our target
is at the edge of the validity range of this index. We also calculated the
KIJ index defined in \citet{2013ApJ...772...79A} obtaining a value of
1.032 that is coherent with an intermediate value for the surface gravity.
From all these results, we can then conclude that HD\,1160\,B is an
intermediate gravity object with a $\log{g}$ of 3.5-4.0~{\it dex}. \par
Also in this case our results are not in agreement with an age of the system
of the order of 120~Myr and a mass of 0.1~\MSun as proposed by
\citet{2019AJ....158...77C}. Indeed, according to the BT-Settl models, an
object with such mass and age would require a $\log{g}$ of the order of
5.0~{\it dex}. To match the values of $\log{g}$ deduced from our analysis, we
would instead need an object at the younger end (10-20~Myr) of the age
range proposed by \citet{2012ApJ...750...53N} and as a consequence, the lower
mass in the range calculated in Table~\ref{t:allmass} (around 20~\MJup).
\par
\begin{figure}
\centering
\includegraphics[width=\columnwidth]{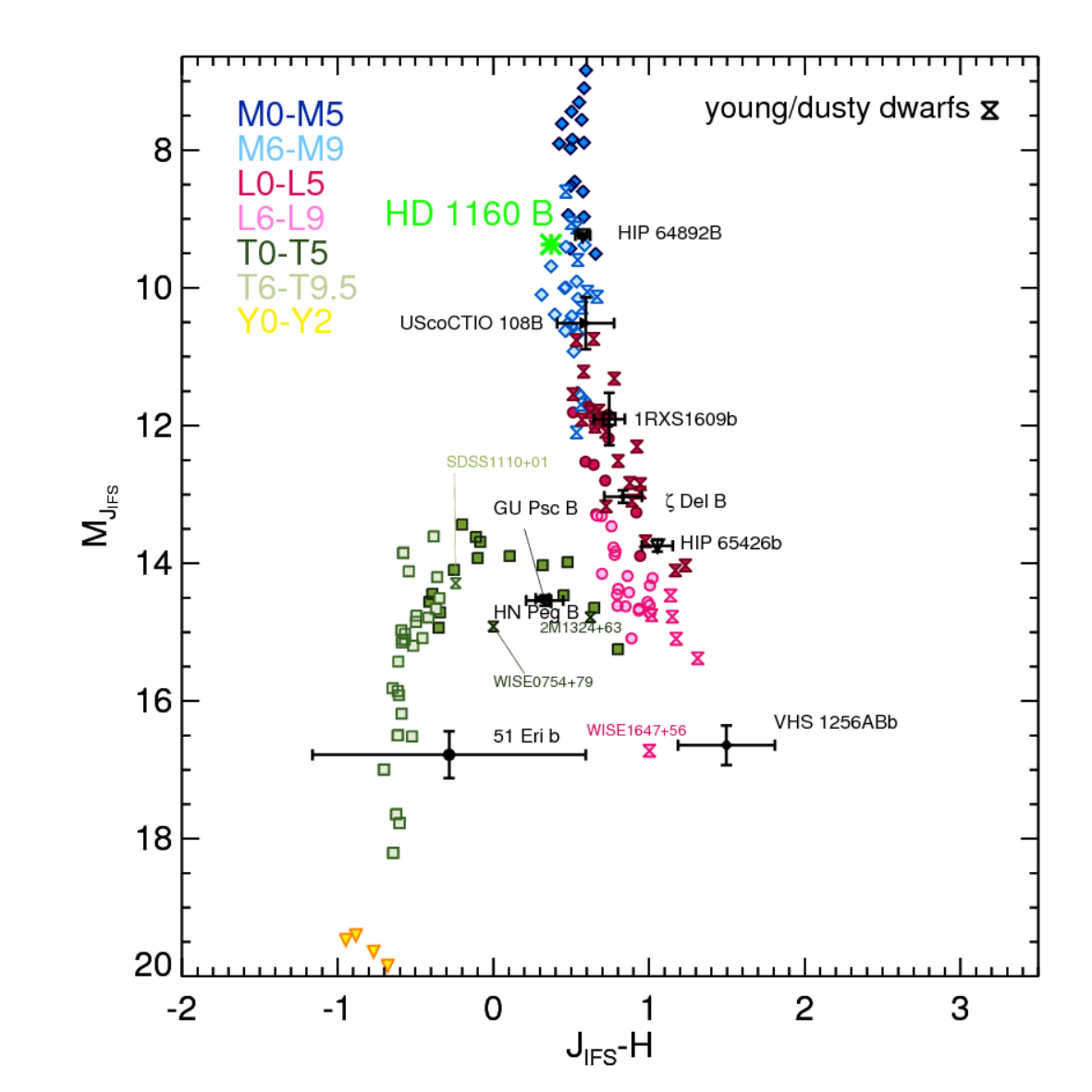}
\caption{Absolute J magnitude vs. J-H color-magnitude diagram with the
  position of HD\,1160\,B that is indicated by a green asterisk. The positions
  of field M, L and T dwarfs are also represented in different colors according
  to the spectral type. Black asterisk represents the positions of other
  substellar companions.}
\label{f:HD1160cmd}
\end{figure}

Using the wide band photometry listed in Table~\ref{t:allphoto} we produced,
following the procedure described in \citet{2018A&A...618A..63B}, the
absolute J magnitude vs J-H
color-magnitude diagram that is displayed in Figure~\ref{f:HD1160cmd}. The
position of HD\,1160\,B, indicated by a green asterisk, is compared to those
of M, L and T field dwarfs indicated in different colors according to the
spectral type and to other substellar companions indicated by black
asterisks. Its position is coherent with that of medium M spectral type
objects confirming what we found with previous analysis.

\subsection{HD\,19467\,B}
\label{s:HD19467res}

\begin{figure}
\centering
\includegraphics[width=\columnwidth]{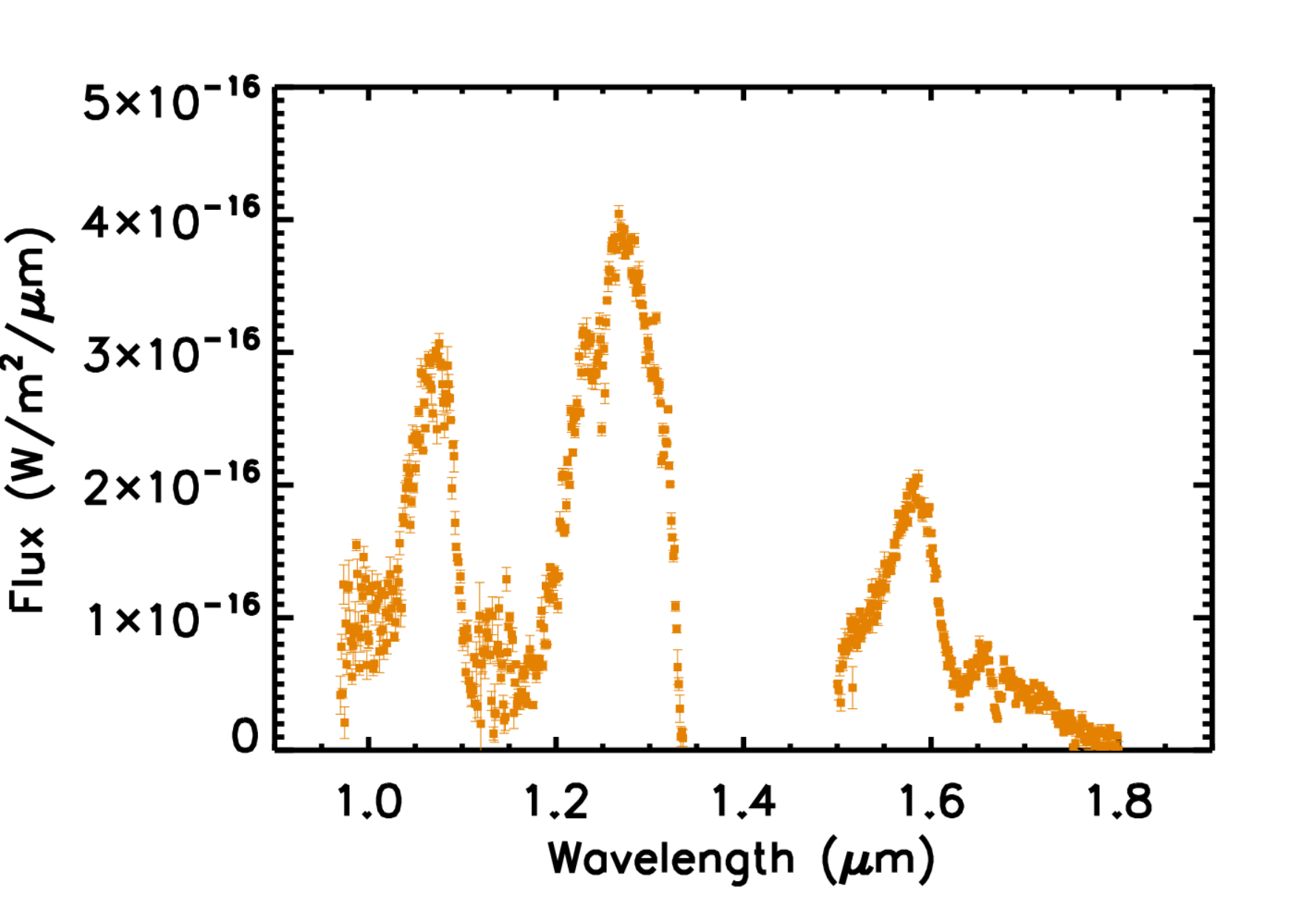}
\caption{Extracted spectrum for HD\,19467\,B using the method described
  in Section~\ref{s:HD19467res}.}
\label{f:HD19467spec}
\end{figure}

Following the same procedure used for HD\,1160\,B we transformed the contrast
spectrum of HD\,19467\,B into the flux spectrum displayed in
Figure~\ref{f:HD19467spec}. In this case the BT-Nextgen synthetic model with
the best fit to the stellar SED and used for this procedure was that with
$T_{eff}$=5600~K, a surface gravity with $\log{g}$=4.5~{\it dex} and solar
metallicity. As for HD\,1160\,B we also used the extracted spectrum to
calculate the absolute magnitude and to estimate the mass of the companion. To
this aim we adopted the system age of 8.0$^{+2.0}_{-1}$ Gyr as mentioned in
Section~\ref{s:target}. Because the magnitude of HD\,19467\,B is outside the
range covered by the BT-Settl models, we employed the AMES-Cond grid of
atmospheric models \citep{2003IAUS..211..325A}. The values resulting from this
procedure are listed in the third column of Table~\ref{t:allphoto} and in the
fourth column of Table~\ref{t:allmass}. In this case the uncertainties are
small and the mass ranges between 60 and 68~\MJup, compatible to what
found by \citet{2019ApJ...873...83W}.

\begin{figure}
\centering
\includegraphics[width=\columnwidth]{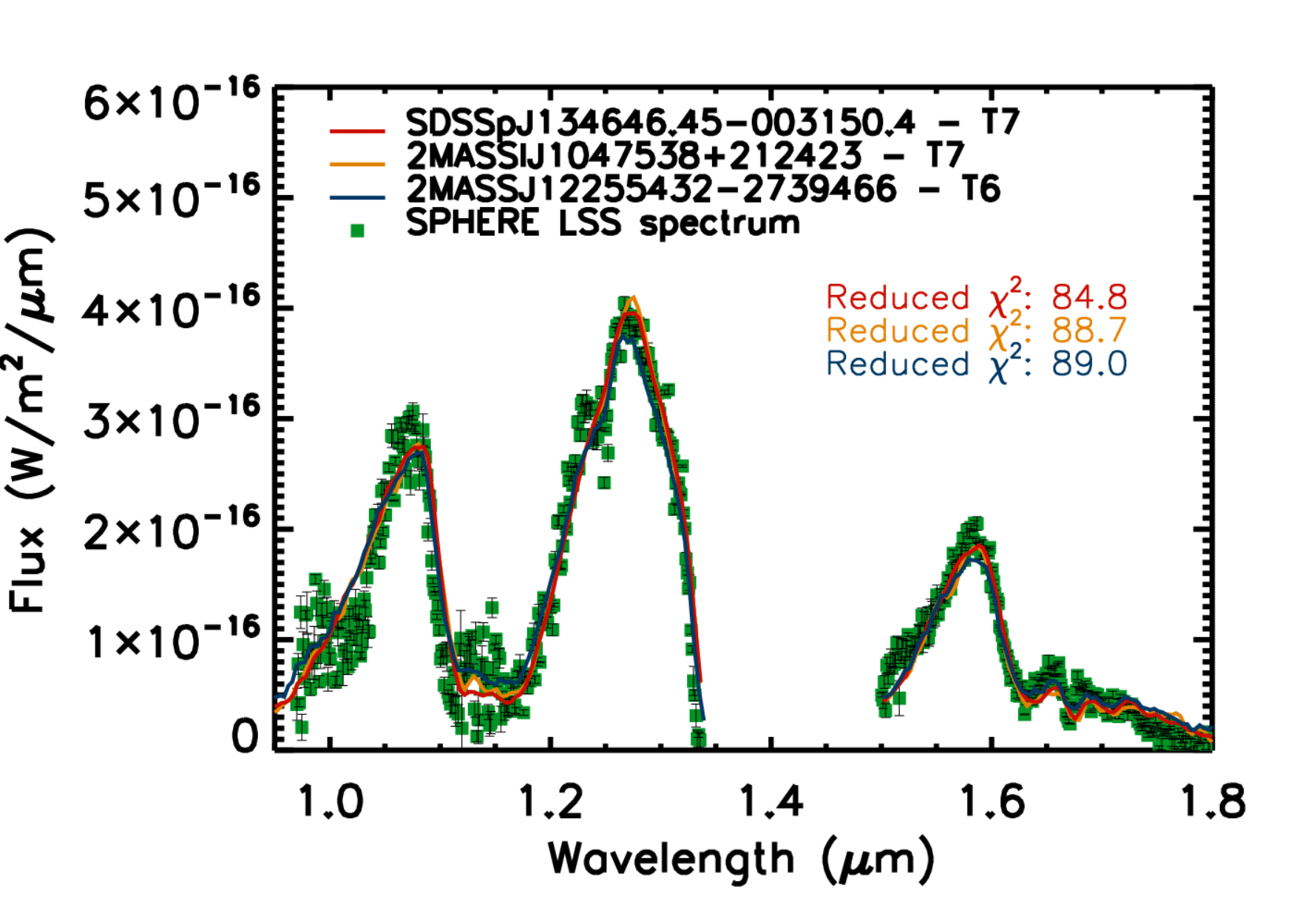}
\caption{Comparison of the extracted spectrum for HD\,19467\,B (green squares)
  with the three best fit spectra from the {\it SpexPrism} library (red,
  orange and blue solid lines).}
\label{f:HD19467spt}
\end{figure}

We followed for HD\,19467\,B the same procedure used for HD\,1160\,B
comparing its spectrum with those
of template spectra. In this case, however, we were not able to find a large
library of high resolution spectra in the T spectral type regime. For this
reason we used the lower resolution spectra in the {\it Spex Prism spectral libraries}\footnote{\url{http://pono.ucsd.edu/~adam/browndwarfs/spexprism/}}
\citep{2014ASInC..11....7B}. This will
not allow to exploit the full potential of the spectral lines in the spectral
classification but will however allow to perform a reliable definition of the
spectral type. The results of this procedure
are displayed in Figure~\ref{f:HD19467spt} where the LSS extracted spectrum
is compared to the three best fit spectra. In this case the best fit
is obtained with the T7 spectrum of SDSSp\,J134646.45-003150.4
\citep{2006ApJ...639.1095B} but a comparable good fit is obtained for spectra
with spectral type between T5 and T7. For this reason we can conclude that
the spectral type for this object is T6$\pm$1. \par
As done for HD\,1160\,B we have tried to confirm this spectral classification
calculating appropriate spectral indices. To this aim we have used the $H_2O$
index at 1.15~\mic, the $H_2O$ index at 1.5~\mic and the $CH_4$ index at
1.6~\mic defined by \citet{2002ApJ...564..466G}. The first index is accurate
only for subtyping in the T sequence, the second one is reliable both on the
L and T sequences while the third one is again effective for the T sequence.
We found for these three indices values of 6.35, 5.56 and 3.13, respectively
for the case of HD\,19467\,B. According to the values for the L-T dwarfs
  subtyping listed in Table~5 of \citet{2002ApJ...564..466G}, these values
are coherent with a T6 spectral type. This is in good agreement with
the spectral type defined through the comparison with template spectra. \par
\begin{figure}
\centering
\includegraphics[width=\columnwidth]{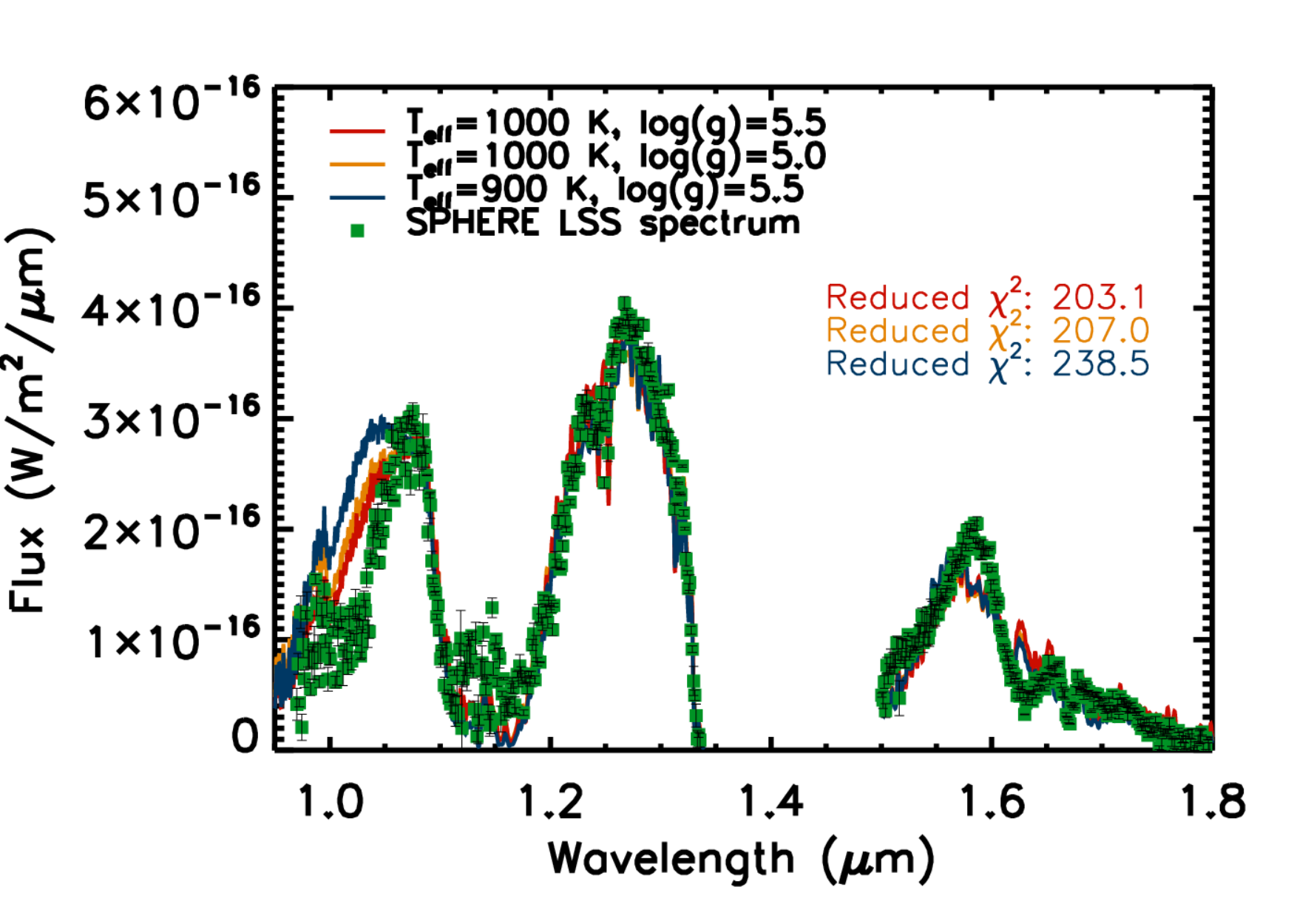}
\caption{Comparison of the extracted spectrum for HD\,19467\,B (green squares)
  with the three best fit theoretical BT-Settl spectra (orange, red and
  blue solid lines).}
\label{f:HD19467synt}
\end{figure}

Similarly to the case of HD\,1160\,B we have fitted also the extracted
spectrum of HD\,19467\,B with a grid of BT-Settl atmospheric models. In this
case their $T_{eff}$ ranges between 500 and 2500~K with a step of 100~K, the
surface gravity ranges between 2.5 and 5.5 {\it dex} while we used solar
metallicity models. The results of this procedure are displayed in
Figure~\ref{f:HD19467synt} where we compare the extracted spectrum with the
three best fit models. The best fit model is obtained for a $T_{eff}$ of 1000~K
and a $\log{g}$ of 5.5~{\it dex}. Good fits were however obtained also for
$T_{eff}$ between 900 and 1100~K while the surface gravity is generally higher
than 4.5~{\it dex}. From these results we can conclude that HD\,19467\,B has
$T_{eff}$=1000$\pm$100~K and $\log{g}$=5.0$\pm$0.5~{\it dex}. In this case the
$T_{eff}$ obtained is in good agreement with what foreseen by
\citet{2016MNRAS.461..794P} for the spectral type that we derived previously
for this object. \par
\begin{figure}
\centering
\includegraphics[width=\columnwidth]{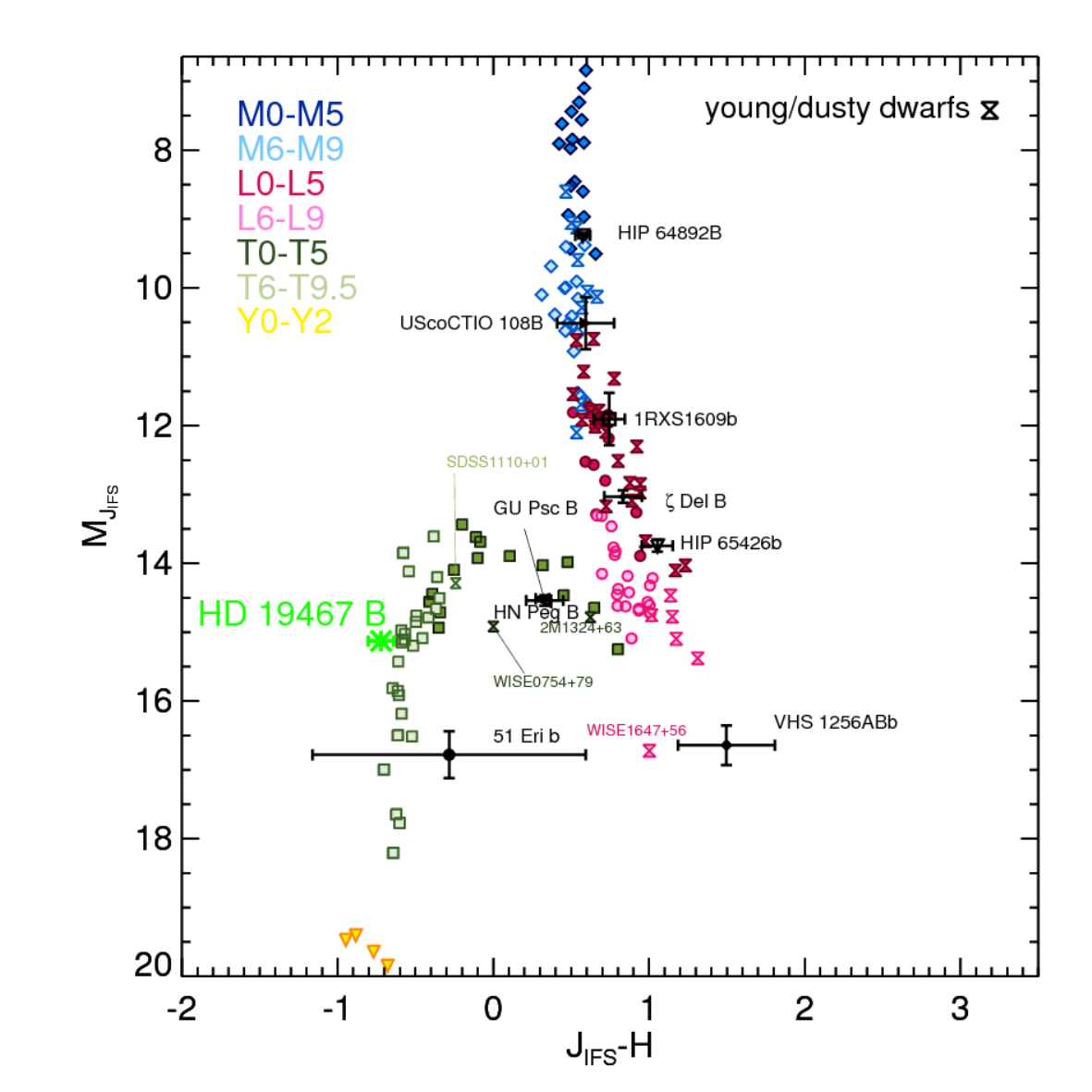}
\caption{Absolute J magnitude vs. J-H color-magnitude diagram with the
  position of HD\,19467\,B that is indicated by a green asterisk. The positions
  of field M, L and T dwarfs are also represented in different colors according
  to the spectral type. Black asterisks represent the positions of other
  substellar companions.}
\label{f:HD19467cmd}
\end{figure}

Following the procedure used for HD\,1160\,B we have created also for
HD\,19467\,B a absolute J magnitude vs J-H color-magnitude diagram (see
Figure~\ref{f:HD19467cmd}). In this diagram, the position of this object is in
good agreement with the expected positions of medium T spectral type objects.

\subsubsection{Atmospheric retrieval of HD\,19467\,B}

In this section we use the extracted spectrum of HD\,19467\,B to perform a retrieval analysis in order to obtain a more in-depth insight into the physical and chemical properties of HD\,19467\,B. For the retrieval, we use the Bayesian, nested-sampling retrieval model \texttt{Helios-r2} \citep{Kitzmann_Helios-r2_09}\footnote{\texttt{Helios-r2} is available as open-source software on the Exoclimes Simulation Platform (ESP) (\url{http://www.exoclime.org/}).}. The model and its application to emission spectra of brown dwarfs is extensively discussed in \citet{Kitzmann2020ApJ...890..174K}. 

Besides the surface gravity $\log g$, the calibration factor $f$, and the distance $d$ we retrieve the abundances of \ch{H2O}, \ch{CH4}, K, and Na. We also tried to include the mixing ratios of \ch{NH3}, \ch{CO}, and \ch{CO2}, which led to unconstrained, prior-dominated posterior distributions. We, therefore, omit these molecules in the final analysis. For the distance $d$ we use a Gaussian prior with a mean value of 32.02 pc and a standard deviation of 0.04 pc, based on GAIA parallax measurements. The temperature profile is described by six first-order elements \citep[see ][ for details on the parametrisation of the temperature profile]{Kitzmann2020ApJ...890..174K}. We also add a grey cloud layer to the retrieval, determined by three parameters (cloud top pressure, the bottom of the
cloud layer, defined as a multiplicative factor of the cloud top pressure, and the grey optical depth). The a priori brown dwarf radius is assumed to be 1~\RJup. Any deviations from this assumed radius are contained in the calibration factor $f$.
In total, we used 18 free parameters, including the parameter aimed to the
inflation of the observational errors following the method described by
\citet{Line2015ApJ...807..183L} and  \citet{Kitzmann2020ApJ...890..174K}.
The inflation of the errors is used to take into account the fact that the
simplified physical model used here is usually not able to describe all the
details in a measured brown dwarf spectrum.

Note that for this retrieval we cut the spectrum below 1.2 $\mu$m. This is necessary to obtain physically consistent values for the surface gravity and the calibration factor. The direct retrieval of the surface gravity and the brown dwarf radius from medium-resolution emission spectra is known to be problematic \citep[e.g. ][]{Oreshenko2020AJ....159....6O, Kitzmann2020ApJ...890..174K}. The spectrum around 1 $\mu$m, which is dominated by the alkali resonance line wings, plays an important role in the determination of $\log g$. Inaccuracies in the measured data, but also the theoretical description of the line wings can strongly influence the retrieved values of the surface gravity. 
We, therefore, did not use this problematic region for the spectral retrieval. However, for comparison, we add a retrieval using the entire available spectral range in the Appendix.

\begin{figure*}
\centering
\includegraphics[width=0.8\textwidth]{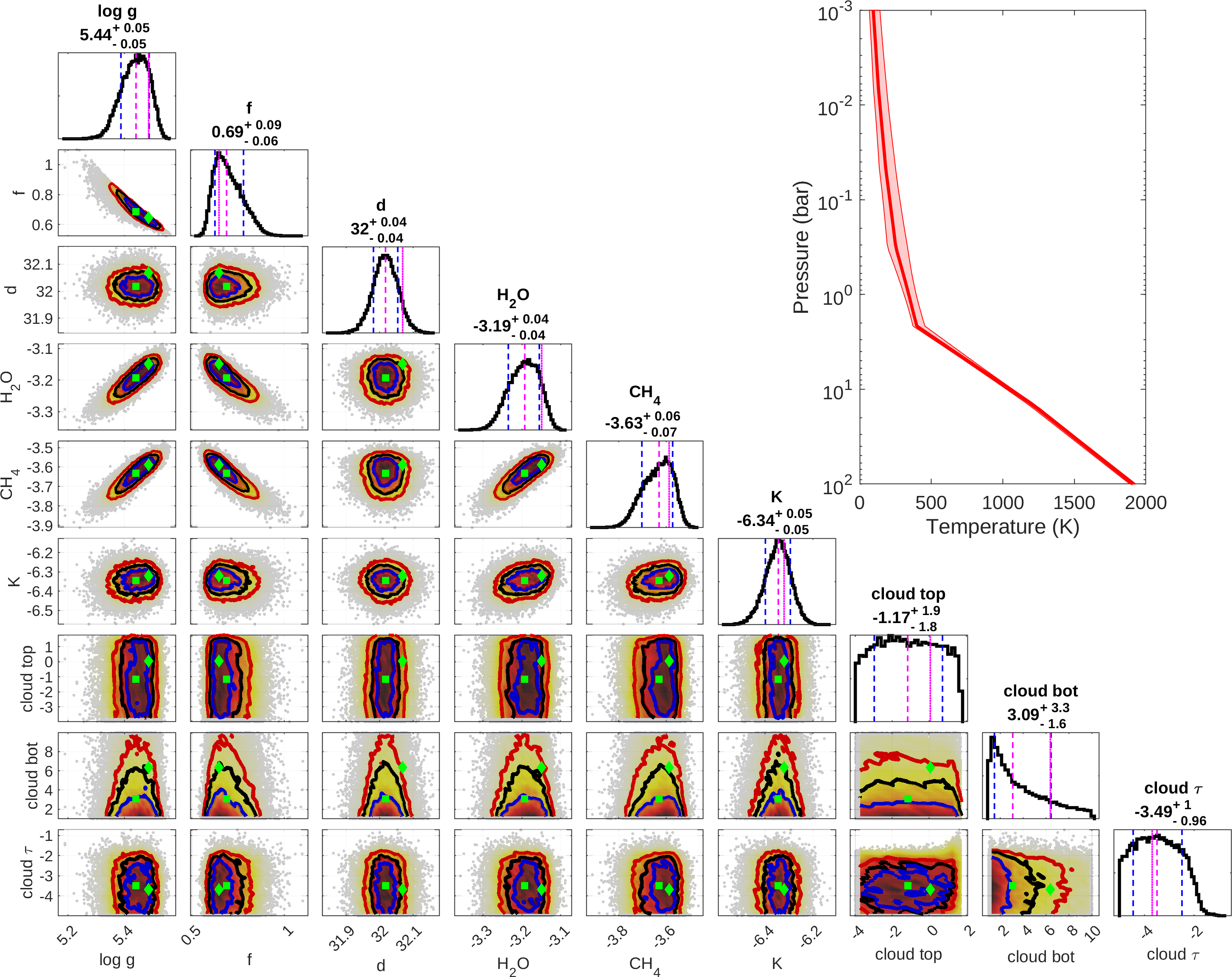}
\caption{Posterior distributions for the retrieval of HD\,19467\,B. The dashed magenta-colored lines in the posterior plots refer to the location of the median value (also stated below each parameter), while the 1$\sigma$ confidence limit is denoted by the blue dashed lines. The magenta dotted line shows the location of the best-fit model, i.e., the one with highest likelihood value. The solid blue, red, and yellow lines in the two-dimensional parameter correlation plots mark the 1$\sigma$, 2$\sigma$, and 3$\sigma$ intervals, respectively. Here, the location of the median (best-fit) model is marked by green squares (diamonds). The panel in the upper, right corner depicts the retrieved temperature profile. The solid red line corresponds to the median profile, while the shaded, red area corresponds to the 1$\sigma$ confidence interval.}
\label{f:HD19467post}
\end{figure*}

\begin{figure}
\centering
\includegraphics[width=\columnwidth]{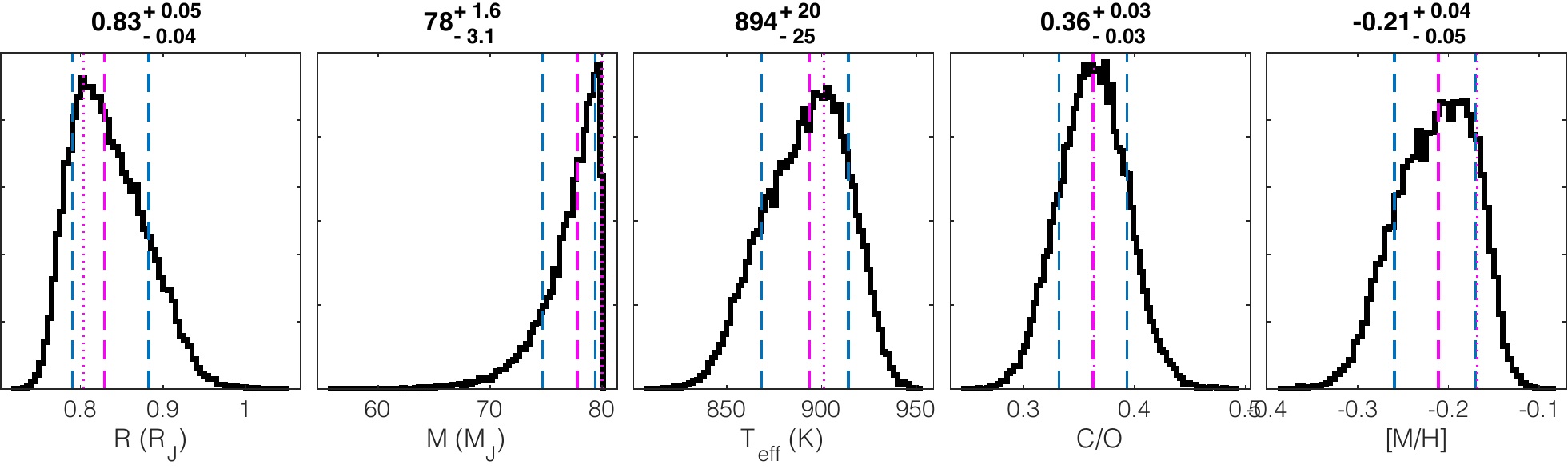}
\includegraphics[width=\columnwidth]{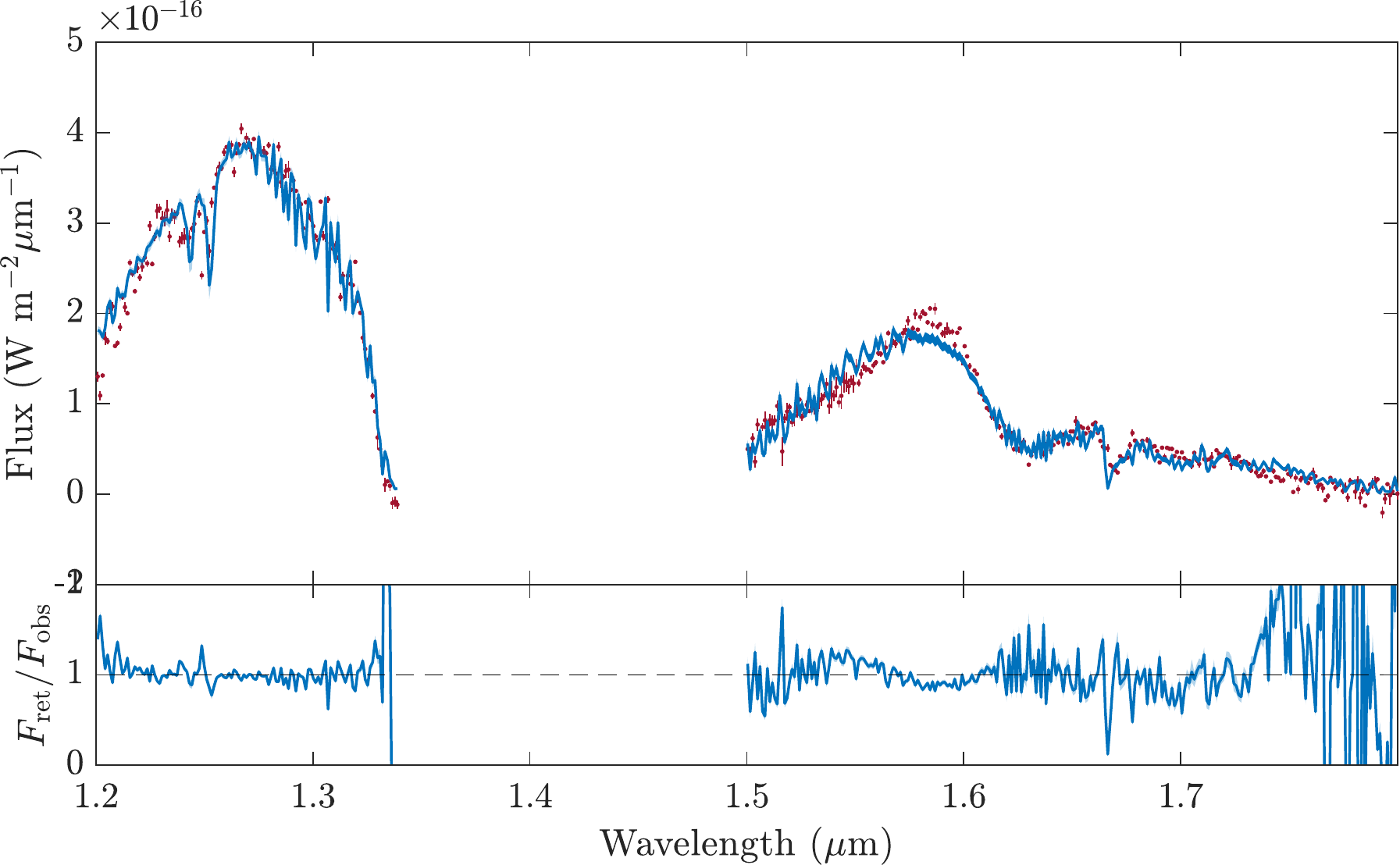}
\caption{{\it Top}: Posterior distributions of derived quantities for the retrieval of HD\,19467\,B. The posteriors are shown for the brown dwarf radius $R$, its mass $M$, effective temperature $T_\mathrm{eff}$, the C/O ratio, and the overall metallicity [M/H]. {\it Bottom}: Posterior spectra and residuals. The solid lines refer to the median of all posterior spectra. The shaded areas signify the 3$\sigma$ confidence intervals of the spectra. The measured spectrum is indicated by the red data points.}
\label{f:HD19467post_add}
\end{figure}

The resulting posterior distributions are shown in Figure \ref{f:HD19467post}, while derived quantities and the posterior spectra are presented in Figure~\ref{f:HD19467post_add}. The retrieved values for the surface gravity $\log g$ are quite high, while the calibration factor $f$ is smaller than unity. On the other hand, the obtained value of $\log g = 5.44 \pm 0.05$~{\it dex} is within the confidence interval of the estimates based on the BT-Settl models, where a value of $5\pm0.5$~{\it dex} is found (see Section \ref{s:HD19467res}).
By assuming that $f$ only describes deviations with respect to assumed radius of 1~\RJup, it can be converted into a posterior for the the brown dwarf radius, shown in Figure \ref{f:HD19467post_add}. In reality, however, $f$ also includes other sources of errors, for example, inaccuracies in the calibration of the spectrum, systematic errors of the instruments and detector, or errors in the distance measurements. 
The retrieved value for $f$ of $0.69 \pm 0.09$ results in a radius estimate of $0.83\pm0.06$~\RJup, which is consistent with evolutionary calculations by \citet{2008ApJ...689.1327S} for such an old object.

The retrieved abundances of \ch{H2O}, \ch{CH4}, and \ch{K} are all well constrained. As mentioned above, adding \ch{NH3}, \ch{CO}, and \ch{CO2} as free parameters will result in unconstrained, prior-dominated posterior distributions. Based on the retrieved abundances, we estimate the C/O ratio and metallicity in Figure~\ref{f:HD19467post_add}. Both, the C/O ratio and the overall metallicity are sub-solar. Within its confidence intervals, this is consistent with the values
found for the host star (Maire et al, submitted).
We note, however, that both quantities are only estimates based on the \ch{H2O}, \ch{CH4}, and \ch{K} values (\ch{H2} and He are assumed to form the background of the atmosphere) and that other molecules that are not detected in the available part of the spectrum might change these estimates.

All cloud parameters are essentially unconstrained which suggests that the presence of a grey cloud layer is not favoured by the data. The temperature profile, on the other hand, is well constrained as evidenced by the small
  uncertainties shown by the shaded area in the right panel of
  Figure~\ref{f:HD19467post}.

The effective temperature presented in Figure~\ref{f:HD19467post_add} has been obtained by integrating the posterior spectra over wavelengths and converting the flux to an effective temperature using the Stefan-Boltzmann law. Its value of about 900 K agrees well with the estimates derived from BT-Settl models (see Section~\ref{s:HD19467res}). 

Due to the small radius and high surface gravity, the derived mass shown in Figure~\ref{f:HD19467post_add} is also quite large. The obtained value of $78^{+1.6}_{-3.1}\,M_\mathrm{J}$ is almost at the edge of the usually assumed upper mass limit for brown dwarfs ($80\,M_\mathrm{J}$). As can be seen in Figure~\ref{f:HD19467post_add}, the posterior distribution is also cut at higher masses. Within the retrieval, we reject all masses larger than $80\, M_\mathrm{J}$ to not allow the
nested sampling to move into an unrealistic mass regime. Without that cut-off, the retrieved mass would even be a bit higher than the value we obtain here.

The posterior spectra shown on the bottom panel of Figure~\ref{f:HD19467post_add} suggest that overall the retrieval matches the measured spectrum quite well. Notable discrepancies are found in the red part of the spectrum. Here, we underestimate the peak in the flux near 1.55 $\mu$m, whereas at the red edge, we slightly overestimate the continuum flux. This might be caused by an additional molecular absorber that has not been included in the retrieval.

\section{Summary and concluding remarks}
\label{s:conclusion}
We presented in this paper SPHERE LSS mode observations of the two benchmark brown-dwarf companions HD\,1160\,B and HD\,19467\,B. The high-quality (SNR and resolution) spectra for these
two objects allowed us to obtain a precise spectral calibration and to gather information on their main physical properties. \par
For the case of HD\,1160\,B the main source of the errors on its physical parameters is due to the large uncertainty on its age. This is important in the derivation of its mass. We have assumed here a range between 10 and 100~Myr that leads to a mass ranging between low-mass brown dwarfs to the low-mass stars regime.
We also tested an older age of $\sim$120~Myr, recently proposed for the
HD\,1160 system that results in a low-star mass for HD\,1160\,B.
The comparison with template spectra gives the best fit with M6$\pm$1 spectral
types. Anyway our procedure is not able to obtain satisfactory fits both in
the Y+J and the H spectral band, simultaneously. Indeed, performing a fit
separately in the Y+J and the H spectral bands tends to favour later and
earlier spectral types, respectively. This dichotomy was also confirmed by the
two spectral indices linked to the $H_2O$ molecule, the first in the J-band and
the second in the H-band. We conclude that the HD\,1160\,B spectrum is very
peculiar given that it is not possible to find a similar spectrum in the
currently available spectral libraries. This could be due to a young age of the
system and/or to the presence of dust in the photosphere of this object. \par
We also compared the spectrum to BT-Settl atmospheric models and we found that
the best fit is obtained for models with $T_{eff}$ of 2800-2900~K and
a surface gravity of $\log{g}$=3.5-4.0~{\it dex}. Using only the spectral
region between 1.10 and 1.30~\mic that contains important alkali (Na~{\sc i} and
K~{\sc i}) lines that are good indicators of the surface gravity allows us to
further confirm the value found previously. Moreover, an intermediate value for
the surface gravity is also confirmed using apposite spectral indices. \par
This last result has an important consequence on the age (and then on the mass)
of HD\,1160\,B. Indeed according to the atmospheric models, a stellar mass
for this object, implied if we assume the older ages (120~Myr) proposed
for the system, would require a surface gravity of
the order of 5.0~{\it dex} that seems to be ruled out by our analysis.
Instead, the surface gravity that we have defined for this object favour
the younger ages (10-20~Myr) and lower masses (around 20~\MJup) in the
proposed ranges. In any case, uncertainties on the fit with the models
and in the determination of the spectral indices do not allow to
fully exclude the possibility of an older age. Further analysis at
higher spectral resolution and with different instruments will be
necessary to completely solve these uncertainties. \par
HD\,19467\,B is instead an old object and its mass can be defined with small uncertainties confirming that it is at the edge between the brown dwarfs and the low stellar mass regimes. We derived a T6 spectral type, through the comparison with templates and using apposite spectral indices. This is in good agreement with the $T_{eff}$ of the order of $\approx$1000~K
that we found from the comparison of its spectrum with BT-Settl atmospheric
models. Finally, we derived a large value for the surface gravity
($\log{g}$=5.0~{\it dex}) that confirms the old age of this fully evolved
object. \par
Finally, we performed on the latter object spectral retrieval to further define
its physical and chemical characteristics. $T_{eff}$ and surface gravity were
in good agreement with what we found using the evolutive models while the
mass of the object tends to be higher than what we found with the models.
From the point of view of the chemical abundances we were able to put strong
constraints on the abundances of \ch{H2O}, \ch{CH4}, and \ch{K} while this
was not possible for other molecules like \ch{NH3}, \ch{CO}, and \ch{CO2}.
Furtermore, our analysis do not favour the presence in the atmosphere of
this object of a grey cloud layer. \par
It was not possible to apply the spectral retrieval analysis to the case of
HD\,1160\,B due to the fact that these models are at the moment optimized to
the case of cold objects and require to be adapted for the case of higher
$T_{eff}$. We defer this to a future work.

\section*{Acknowledgments}
SPHERE is an instrument designed and built by a consortium
consisting of IPAG (Grenoble, France), MPIA (Heidelberg, Germany), LAM
(Marseille, France), LESIA (Paris, France), Laboratoire Lagrange
(Nice, France), INAF–Osservatorio di Padova (Italy), Observatoire de
Genève (Switzerland), ETH Zurich (Switzerland), NOVA (Netherlands),
ONERA (France) and ASTRON (Netherlands) in collaboration with
ESO. SPHERE was funded by ESO, with additional contributions from CNRS
(France), MPIA (Germany), INAF (Italy), FINES (Switzerland) and NOVA
(Netherlands).  SPHERE also received funding from the European
Commission Sixth and Seventh Framework Programmes as part of the
Optical Infrared Coordination Network for Astronomy (OPTICON) under
grant number RII3-Ct-2004-001566 for FP6 (2004–2008), grant number
226604 for FP7 (2009–2012) and grant number 312430 for FP7
(2013–2016). We also acknowledge financial support from the Programme National
de Plan\'{e}tologie (PNP) and the Programme National de Physique Stellaire
(PNPS) of CNRS-INSU in France. This work has also been supported by a grant from
the French Labex OSUG@2020 (Investissements d’avenir – ANR10 LABX56).
The project is supported by CNRS, by the Agence Nationale de la
Recherche (ANR-14-CE33-0018). It has also been carried out within the frame of
the National Centre for Competence in  Research PlanetS supported by the Swiss
National Science Foundation (SNSF). \par
This research has made use of the SIMBAD database, operated at CDS,
Strasbourg, France. \par
This research has benefited from the Montreal Brown Dwarf and Exoplanet Spectral
Library, mantained by Jonathan Gagn'{e}. \par
This publication makes use of VOSA, developed under the Spanish Virtual Observatory project supported by the Spanish MINECO through grant AyA2017-84089.
VOSA has been partially updated by using funding from the European Union's Horizon 2020 Research and Innovation Programme, under Grant Agreement nº 776403 (EXOPLANETS-A). \par
D.M., V.D.O., R.G., S.D. acknowledge support from
the ``Progetti Premiali'' funding scheme of the Italian Ministry of Education,
University, and Research. \par
D.K. acknowledges partial financial support from the Center for Space and Habitability (CSH) of the University of Bern and the PlanetS National Center of Competence in Research (NCCR).

\bibliographystyle{mn2e}
\bibliography{LSS}

\begin{thebibliography}{}

\bibitem[\protect\citeauthoryear{{Allard}}{{Allard}}{2014}]{2014IAUS..299..271A}
{Allard} F.,  2014, in {Booth} M.,  {Matthews} B.~C.,   {Graham} J.~R.,  eds,
  Exploring the Formation and Evolution of Planetary Systems Vol.~299 of IAU
  Symposium, {The BT-Settl Model Atmospheres for Stars, Brown Dwarfs and
  Planets}.
pp 271--272

\bibitem[\protect\citeauthoryear{{Allard}, {Guillot}, {Ludwig} \& et
  al.}{{Allard} et~al.}{2003}]{2003IAUS..211..325A}
{Allard} F.,  {Guillot} T.,  {Ludwig} H.-G.,    et al. 2003, in {Mart{\'{\i}}n}
  E.,  ed., Brown Dwarfs Vol.~211 of IAU Symposium, {Model Atmospheres and
  Spectra: The Role of Dust}.
p.~325

\bibitem[\protect\citeauthoryear{{Allard}, {Homeier} \& {Freytag}}{{Allard}
  et~al.}{2012}]{2012RSPTA.370.2765A}
{Allard} F.,  {Homeier} D.,    {Freytag} B.,  2012, Philosophical Transactions
  of the Royal Society of London Series A, 370, 2765

\bibitem[\protect\citeauthoryear{{Allers}, {Jaffe}, {Luhman} \& et
  al.}{{Allers} et~al.}{2007}]{2007ApJ...657..511A}
{Allers} K.~N.,  {Jaffe} D.~T.,  {Luhman} K.~L.,    et al. 2007, \apj, 657, 511

\bibitem[\protect\citeauthoryear{{Allers} \& {Liu}}{{Allers} \&
  {Liu}}{2013}]{2013ApJ...772...79A}
{Allers} K.~N.,  {Liu} M.~C.,  2013, \apj, 772, 79

\bibitem[\protect\citeauthoryear{{Allers}, {Liu}, {Dupuy} \&
  {Cushing}}{{Allers} et~al.}{2010}]{2010ApJ...715..561A}
{Allers} K.~N.,  {Liu} M.~C.,  {Dupuy} T.~J.,    {Cushing} M.~C.,  2010, \apj,
  715, 561

\bibitem[\protect\citeauthoryear{{Baraffe}, {Chabrier}, {Barman} \& et
  al.}{{Baraffe} et~al.}{2003}]{baraffe2003}
{Baraffe} I.,  {Chabrier} G.,  {Barman} T.~S.,    et al. 2003, \aap, 402, 701

\bibitem[\protect\citeauthoryear{{Bayo}, {Rodrigo}, {Barrado Y Navascu{\'e}s}
  \& et al.}{{Bayo} et~al.}{2008}]{2008A&A...492..277B}
{Bayo} A.,  {Rodrigo} C.,  {Barrado Y Navascu{\'e}s} D.,    et al. 2008, \aap,
  492, 277

\bibitem[\protect\citeauthoryear{{Bensby}, {Feltzing} \& {Oey}}{{Bensby}
  et~al.}{2014}]{bensby2014}
{Bensby} T.,  {Feltzing} S.,    {Oey} M.~S.,  2014, \aap, 562, A71

\bibitem[\protect\citeauthoryear{{Beuzit}, {Vigan}, {Mouillet} \& et
  al.}{{Beuzit} et~al.}{2019}]{2019A&A...631A.155B}
{Beuzit} J.~L.,  {Vigan} A.,  {Mouillet} D.,    et al. 2019, \aap, 631, A155

\bibitem[\protect\citeauthoryear{{Bonavita}, {D'Orazi}, {Mesa}, {Fontanive} \&
  et al.}{{Bonavita} et~al.}{2017}]{2017A&A...608A.106B}
{Bonavita} M.,  {D'Orazi} V.,  {Mesa} D.,  {Fontanive} C.,    et al. 2017,
  \aap, 608, A106

\bibitem[\protect\citeauthoryear{{Bonnefoy}, {Chauvin}, {Lagrange} \& et
  al.}{{Bonnefoy} et~al.}{2014}]{2014A&A...562A.127B}
{Bonnefoy} M.,  {Chauvin} G.,  {Lagrange} A.~M.,    et al. 2014, \aap, 562,
  A127

\bibitem[\protect\citeauthoryear{{Bonnefoy}, {Perraut}, {Lagrange} \& et
  al.}{{Bonnefoy} et~al.}{2018}]{2018A&A...618A..63B}
{Bonnefoy} M.,  {Perraut} K.,  {Lagrange} A.~M.,    et al. 2018, \aap, 618, A63

\bibitem[\protect\citeauthoryear{{Burgasser}}{{Burgasser}}{2014}]{2014ASInC..11....7B}
{Burgasser} A.~J.,  2014, in Astronomical Society of India Conference Series
  Vol.~11 of Astronomical Society of India Conference Series, {The SpeX Prism
  Library: 1000+ low-resolution, near-infrared spectra of ultracool M, L, T and
  Y dwarfs}.
pp 7--16

\bibitem[\protect\citeauthoryear{{Burgasser}, {Burrows} \&
  {Kirkpatrick}}{{Burgasser} et~al.}{2006}]{2006ApJ...639.1095B}
{Burgasser} A.~J.,  {Burrows} A.,    {Kirkpatrick} J.~D.,  2006, \apj, 639,
  1095

\bibitem[\protect\citeauthoryear{{Cameron}}{{Cameron}}{1978}]{1978M&P....18....5C}
{Cameron} A.~G.~W.,  1978, Moon and Planets, 18, 5

\bibitem[\protect\citeauthoryear{{Cheetham}, {Bonnefoy}, {Desidera} \& et
  al.}{{Cheetham} et~al.}{2018}]{2018A&A...615A.160C}
{Cheetham} A.,  {Bonnefoy} M.,  {Desidera} S.,    et al. 2018, \aap, 615, A160

\bibitem[\protect\citeauthoryear{{Chun}, {Toomey}, {Wahhaj} \& et al.}{{Chun}
  et~al.}{2008}]{2008SPIE.7015E..1VC}
{Chun} M.,  {Toomey} D.,  {Wahhaj} Z.,    et al. 2008, {Performance of the
  near-infrared coronagraphic imager on Gemini-South}.
p. 70151V

\bibitem[\protect\citeauthoryear{{Crepp}, {Johnson}, {Howard} \& et
  al.}{{Crepp} et~al.}{2014}]{2014ApJ...781...29C}
{Crepp} J.~R.,  {Johnson} J.~A.,  {Howard} A.~W.,    et al. 2014, \apj, 781, 29

\bibitem[\protect\citeauthoryear{{Crepp}, {Principe}, {Wolff} \& et
  al.}{{Crepp} et~al.}{2018}]{2018ApJ...853..192C}
{Crepp} J.~R.,  {Principe} D.~A.,  {Wolff} S.,    et al. 2018, \apj, 853, 192

\bibitem[\protect\citeauthoryear{{Crepp}, {Rice}, {Veicht} \& et al.}{{Crepp}
  et~al.}{2015}]{2015ApJ...798L..43C}
{Crepp} J.~R.,  {Rice} E.~L.,  {Veicht} A.,    et al. 2015, \apjl, 798, L43

\bibitem[\protect\citeauthoryear{{Cruz}, {Kirkpatrick} \& {Burgasser}}{{Cruz}
  et~al.}{2009}]{2009AJ....137.3345C}
{Cruz} K.~L.,  {Kirkpatrick} J.~D.,    {Burgasser} A.~J.,  2009, \aj, 137, 3345

\bibitem[\protect\citeauthoryear{{Curtis}, {Ag{\"u}eros}, {Mamajek} \& et
  al.}{{Curtis} et~al.}{2019}]{2019AJ....158...77C}
{Curtis} J.~L.,  {Ag{\"u}eros} M.~A.,  {Mamajek} E.~E.,    et al. 2019, \aj,
  158, 77

\bibitem[\protect\citeauthoryear{{Cushing}, {Rayner} \& {Vacca}}{{Cushing}
  et~al.}{2005}]{2005ApJ...623.1115C}
{Cushing} M.~C.,  {Rayner} J.~T.,    {Vacca} W.~D.,  2005, \apj, 623, 1115

\bibitem[\protect\citeauthoryear{{De Marchi}, {Paresce} \& {Portegies
  Zwart}}{{De Marchi} et~al.}{2010}]{2010ApJ...718..105D}
{De Marchi} G.,  {Paresce} F.,    {Portegies Zwart} S.,  2010, \apj, 718, 105

\bibitem[\protect\citeauthoryear{{Delorme}, {Schmidt}, {Bonnefoy} \& et
  al.}{{Delorme} et~al.}{2017}]{2017A&A...608A..79D}
{Delorme} P.,  {Schmidt} T.,  {Bonnefoy} M.,    et al. 2017, \aap, 608, A79

\bibitem[\protect\citeauthoryear{{Dohlen}, {Langlois}, {Saisse} \& et
  al.}{{Dohlen} et~al.}{2008}]{2008SPIE.7014E..3LD}
{Dohlen} K.,  {Langlois} M.,  {Saisse} M.,    et al. 2008, in Ground-based and
  Airborne Instrumentation for Astronomy II Vol.~7014 of \procspie, {The
  infra-red dual imaging and spectrograph for SPHERE: design and performance}.
p. 70143L

\bibitem[\protect\citeauthoryear{{Gagn{\'e}}, {Faherty}, {Cruz} \& et
  al.}{{Gagn{\'e}} et~al.}{2015}]{2015ApJS..219...33G}
{Gagn{\'e}} J.,  {Faherty} J.~K.,  {Cruz} K.~L.,    et al. 2015, \apjs, 219, 33

\bibitem[\protect\citeauthoryear{{Gaia Collaboration}}{{Gaia
  Collaboration}}{2018}]{2018yCat.1345....0G}
{Gaia Collaboration} 2018, VizieR Online Data Catalog, p. I/345

\bibitem[\protect\citeauthoryear{{Garcia}, {Currie}, {Guyon} \& et
  al.}{{Garcia} et~al.}{2017}]{2017ApJ...834..162G}
{Garcia} E.~V.,  {Currie} T.,  {Guyon} O.,    et al. 2017, \apj, 834, 162

\bibitem[\protect\citeauthoryear{{Geballe}, {Knapp}, {Leggett} \& et
  al.}{{Geballe} et~al.}{2002}]{2002ApJ...564..466G}
{Geballe} T.~R.,  {Knapp} G.~R.,  {Leggett} S.~K.,    et al. 2002, \apj, 564,
  466

\bibitem[\protect\citeauthoryear{{Gorlova}, {Meyer}, {Rieke} \&
  {Liebert}}{{Gorlova} et~al.}{2003}]{2003ApJ...593.1074G}
{Gorlova} N.~I.,  {Meyer} M.~R.,  {Rieke} G.~H.,    {Liebert} J.,  2003, \apj,
  593, 1074

\bibitem[\protect\citeauthoryear{{Hayashi} \& {Nakano}}{{Hayashi} \&
  {Nakano}}{1963}]{1963PThPh..30..460H}
{Hayashi} C.,  {Nakano} T.,  1963, Progress of Theoretical Physics, 30, 460

\bibitem[\protect\citeauthoryear{{Hinkley}, {Bowler}, {Vigan} \& et
  al.}{{Hinkley} et~al.}{2015}]{2015ApJ...805L..10H}
{Hinkley} S.,  {Bowler} B.~P.,  {Vigan} A.,    et al. 2015, \apjl, 805, L10

\bibitem[\protect\citeauthoryear{{Hodapp}, {Jensen}, {Irwin} \& et
  al.}{{Hodapp} et~al.}{2003}]{2003PASP..115.1388H}
{Hodapp} K.~W.,  {Jensen} J.~B.,  {Irwin} E.~M.,    et al. 2003, \pasp, 115,
  1388

\bibitem[\protect\citeauthoryear{{Houk} \& {Smith-Moore}}{{Houk} \&
  {Smith-Moore}}{1988}]{1988mcts.book.....H}
{Houk} N.,  {Smith-Moore} M.,  1988, {Michigan Catalogue of Two-dimensional
  Spectral Types for the HD Stars. Volume 4}.
Vol.~4

\bibitem[\protect\citeauthoryear{{Houk} \& {Swift}}{{Houk} \&
  {Swift}}{1999}]{1999MSS...C05....0H}
{Houk} N.,  {Swift} C.,  1999, Michigan Spectral Survey, 5, 0

\bibitem[\protect\citeauthoryear{{Johnson-Groh}, {Marois}, {De Rosa} \& et
  al.}{{Johnson-Groh} et~al.}{2017}]{2017AJ....153..190J}
{Johnson-Groh} M.,  {Marois} C.,  {De Rosa} R.~J.,    et al. 2017, \aj, 153,
  190

\bibitem[\protect\citeauthoryear{{Kiefer}, {H{\'e}brard}, {Sahlmann} \& et
  al.}{{Kiefer} et~al.}{2019}]{2019A&A...631A.125K}
{Kiefer} F.,  {H{\'e}brard} G.,  {Sahlmann} J.,    et al. 2019, \aap, 631, A125

\bibitem[\protect\citeauthoryear{{Kirkpatrick}}{{Kirkpatrick}}{2005}]{2005ARA&A..43..195K}
{Kirkpatrick} J.~D.,  2005, \araa, 43, 195

\bibitem[\protect\citeauthoryear{{Kitzmann}}{{Kitzmann}}{2020}]{Kitzmann_Helios-r2_09}
{Kitzmann} D., , 2020, {Helios-r2 - A Bayesian Nested Sampling Retrieval Code}

\bibitem[\protect\citeauthoryear{{Kitzmann}, {Heng}, {Oreshenko} \& et
  al.}{{Kitzmann} et~al.}{2020}]{Kitzmann2020ApJ...890..174K}
{Kitzmann} D.,  {Heng} K.,  {Oreshenko} M.,    et al. 2020, \apj, 890, 174

\bibitem[\protect\citeauthoryear{{Konopacky}, {Rameau}, {Duch{\^e}ne} \& et
  al.}{{Konopacky} et~al.}{2016}]{2016ApJ...829L...4K}
{Konopacky} Q.~M.,  {Rameau} J.,  {Duch{\^e}ne} G.,    et al. 2016, \apjl, 829,
  L4

\bibitem[\protect\citeauthoryear{{Kumar}}{{Kumar}}{1963}]{1963ApJ...137.1126K}
{Kumar} S.~S.,  1963, \apj, 137, 1126

\bibitem[\protect\citeauthoryear{{Lenzen}, {Hartung}, {Brandner} \& et
  al.}{{Lenzen} et~al.}{2003}]{2003SPIE.4841..944L}
{Lenzen} R.,  {Hartung} M.,  {Brandner} W.,    et al. 2003, in {Iye} M.,
  {Moorwood} A.~F.~M.,  eds, Instrument Design and Performance for
  Optical/Infrared Ground-based Telescopes Vol.~4841 of \procspie, {NAOS-CONICA
  first on sky results in a variety of observing modes}.
pp 944--952

\bibitem[\protect\citeauthoryear{{Line}, {Teske}, {Burningham} \& et
  al.}{{Line} et~al.}{2015}]{Line2015ApJ...807..183L}
{Line} M.~R.,  {Teske} J.,  {Burningham} B.,    et al. 2015, \apj, 807, 183

\bibitem[\protect\citeauthoryear{{Macintosh}, {Gemini Planet Imager instrument
  Team}, {Planet Imager Exoplanet Survey} \& {Observatory}}{{Macintosh}
  et~al.}{2014}]{2014AAS...22322902M}
{Macintosh} B.,  {Gemini Planet Imager instrument Team} {Planet Imager
  Exoplanet Survey} G.,    {Observatory} G.,  2014, in American Astronomical
  Society Meeting Abstracts \#223 Vol.~223 of American Astronomical Society
  Meeting Abstracts, {The Gemini Planet Imager}.
p. 229.02

\bibitem[\protect\citeauthoryear{{Maire}, {Bonnefoy}, {Ginski} \& et
  al.}{{Maire} et~al.}{2016}]{2016A&A...587A..56M}
{Maire} A.~L.,  {Bonnefoy} M.,  {Ginski} C.,    et al. 2016, \aap, 587, A56

\bibitem[\protect\citeauthoryear{{Martin}, {Mace}, {McLean} \& et al.}{{Martin}
  et~al.}{2017}]{2017ApJ...838...73M}
{Martin} E.~C.,  {Mace} G.~N.,  {McLean} I.~S.,    et al. 2017, \apj, 838, 73

\bibitem[\protect\citeauthoryear{{Mawet}, {Choquet}, {Absil} \& et al.}{{Mawet}
  et~al.}{2017}]{2017AJ....153...44M}
{Mawet} D.,  {Choquet} {\'E}.,  {Absil} O.,    et al. 2017, \aj, 153, 44

\bibitem[\protect\citeauthoryear{{Mawet}, {David}, {Bottom} \& et al.}{{Mawet}
  et~al.}{2015}]{2015ApJ...811..103M}
{Mawet} D.,  {David} T.,  {Bottom} M.,    et al. 2015, \apj, 811, 103

\bibitem[\protect\citeauthoryear{{Meingast}, {Alves} \&
  {F{\"u}rnkranz}}{{Meingast} et~al.}{2019}]{2019A&A...622L..13M}
{Meingast} S.,  {Alves} J.,    {F{\"u}rnkranz} V.,  2019, \aap, 622, L13

\bibitem[\protect\citeauthoryear{{Mesa}, {Baudino}, {Charnay} \& et al.}{{Mesa}
  et~al.}{2018}]{2018A&A...612A..92M}
{Mesa} D.,  {Baudino} J.~L.,  {Charnay} B.,    et al. 2018, \aap, 612, A92

\bibitem[\protect\citeauthoryear{{Mesa}, {Vigan}, {D'Orazi} \& et al.}{{Mesa}
  et~al.}{2016}]{2016A&A...593A.119M}
{Mesa} D.,  {Vigan} A.,  {D'Orazi} V.,    et al. 2016, \aap, 593, A119

\bibitem[\protect\citeauthoryear{{Milli}, {Hibon}, {Christiaens} \& et
  al.}{{Milli} et~al.}{2017}]{2017A&A...597L...2M}
{Milli} J.,  {Hibon} P.,  {Christiaens} V.,    et al. 2017, \aap, 597, L2

\bibitem[\protect\citeauthoryear{{Nakajima}, {Oppenheimer}, {Kulkarni} \& et
  al.}{{Nakajima} et~al.}{1995}]{1995Natur.378..463N}
{Nakajima} T.,  {Oppenheimer} B.~R.,  {Kulkarni} S.~R.,    et al. 1995, \nat,
  378, 463

\bibitem[\protect\citeauthoryear{{Nielsen}, {Liu}, {Wahhaj} \& et
  al.}{{Nielsen} et~al.}{2012}]{2012ApJ...750...53N}
{Nielsen} E.~L.,  {Liu} M.~C.,  {Wahhaj} Z.,    et al. 2012, \apj, 750, 53

\bibitem[\protect\citeauthoryear{{Oreshenko}, {Kitzmann}, {M{\'a}rquez-Neila}
  \& et al.}{{Oreshenko} et~al.}{2020}]{Oreshenko2020AJ....159....6O}
{Oreshenko} M.,  {Kitzmann} D.,  {M{\'a}rquez-Neila} P.,    et al. 2020, \aj,
  159, 6

\bibitem[\protect\citeauthoryear{{Pecaut} \& {Mamajek}}{{Pecaut} \&
  {Mamajek}}{2016}]{2016MNRAS.461..794P}
{Pecaut} M.~J.,  {Mamajek} E.~E.,  2016, \mnras, 461, 794

\bibitem[\protect\citeauthoryear{{Peretti}, {S{\'e}gransan}, {Lavie} \& et
  al.}{{Peretti} et~al.}{2019}]{2019A&A...631A.107P}
{Peretti} S.,  {S{\'e}gransan} D.,  {Lavie} B.,    et al. 2019, \aap, 631, A107

\bibitem[\protect\citeauthoryear{{Rebolo}, {Zapatero Osorio} \&
  {Mart{\'\i}n}}{{Rebolo} et~al.}{1995}]{rebolo95}
{Rebolo} R.,  {Zapatero Osorio} M.~R.,    {Mart{\'\i}n} E.~L.,  1995, \nat,
  377, 129

\bibitem[\protect\citeauthoryear{{Rickman}, {S{\'e}gransan}, {Hagelberg} \& et
  al.}{{Rickman} et~al.}{2020}]{2020arXiv200208319R}
{Rickman} E.~L.,  {S{\'e}gransan} D.,  {Hagelberg} J.,    et al. 2020, arXiv
  e-prints, p. arXiv:2002.08319

\bibitem[\protect\citeauthoryear{{R{\"o}ser} \& {Schilbach}}{{R{\"o}ser} \&
  {Schilbach}}{2020}]{2020arXiv200203610R}
{R{\"o}ser} S.,  {Schilbach} E.,  2020, arXiv e-prints, p. arXiv:2002.03610

\bibitem[\protect\citeauthoryear{{Saumon} \& {Marley}}{{Saumon} \&
  {Marley}}{2008}]{2008ApJ...689.1327S}
{Saumon} D.,  {Marley} M.~S.,  2008, \apj, 689, 1327

\bibitem[\protect\citeauthoryear{{Seeliger}, {Neuh{\"a}user} \&
  {Eisenbeiss}}{{Seeliger} et~al.}{2011}]{2011AN....332..821S}
{Seeliger} M.,  {Neuh{\"a}user} R.,    {Eisenbeiss} T.,  2011, Astronomische
  Nachrichten, 332, 821

\bibitem[\protect\citeauthoryear{{Slesnick}, {Hillenbrand} \&
  {Carpenter}}{{Slesnick} et~al.}{2004}]{2004ApJ...610.1045S}
{Slesnick} C.~L.,  {Hillenbrand} L.~A.,    {Carpenter} J.~M.,  2004, \apj, 610,
  1045

\bibitem[\protect\citeauthoryear{{Soummer}, {Pueyo} \& {Larkin}}{{Soummer}
  et~al.}{2012}]{2012ApJ...755L..28S}
{Soummer} R.,  {Pueyo} L.,    {Larkin} J.,  2012, \apjl, 755, L28

\bibitem[\protect\citeauthoryear{{Spiegel}, {Burrows} \& {Milsom}}{{Spiegel}
  et~al.}{2011}]{2011ApJ...727...57S}
{Spiegel} D.~S.,  {Burrows} A.,    {Milsom} J.~A.,  2011, \apj, 727, 57

\bibitem[\protect\citeauthoryear{{Spina}, {Randich}, {Magrini}, {Jeffries},
  {Friel}, {Sacco}, {Pancino}, {Bonito}, {Bravi}, {Franciosini}, {Klutsch},
  {Montes}, {Gilmore}, {Vallenari}, {Bensby} \& {Bragaglia}}{{Spina}
  et~al.}{2017}]{spina2017}
{Spina} L.,  {Randich} S.,  {Magrini} L.,  {Jeffries} R.~D.,  {Friel} E.~D.,
  {Sacco} G.~G.,  {Pancino} E.,  {Bonito} R.,  {Bravi} L.,  {Franciosini} E.,
  {Klutsch} A.,  {Montes} D.,  {Gilmore} G.,  {Vallenari} A.,  {Bensby} T.,
  {Bragaglia} A. e.~a.,  2017, \aap, 601, A70

\bibitem[\protect\citeauthoryear{{V{\'a}zquez-Semadeni}, {Palau},
  {Ballesteros-Paredes} \& et al.}{{V{\'a}zquez-Semadeni}
  et~al.}{2019}]{2019MNRAS.490.3061V}
{V{\'a}zquez-Semadeni} E.,  {Palau} A.,  {Ballesteros-Paredes} J.,    et al.
  2019, \mnras, 490, 3061

\bibitem[\protect\citeauthoryear{{Vigan}}{{Vigan}}{2016}]{2016ascl.soft03001V}
{Vigan} A., , 2016, {SILSS: SPHERE/IRDIS Long-Slit Spectroscopy pipeline}

\bibitem[\protect\citeauthoryear{{Vigan}, {Bonnefoy}, {Ginski} \& et
  al.}{{Vigan} et~al.}{2016}]{2016A&A...587A..55V}
{Vigan} A.,  {Bonnefoy} M.,  {Ginski} C.,    et al. 2016, \aap, 587, A55

\bibitem[\protect\citeauthoryear{{Vigan}, {Langlois}, {Moutou} \&
  {Dohlen}}{{Vigan} et~al.}{2008}]{2008A&A...489.1345V}
{Vigan} A.,  {Langlois} M.,  {Moutou} C.,    {Dohlen} K.,  2008, \aap, 489,
  1345

\bibitem[\protect\citeauthoryear{{Vigan}, {N'Diaye}, {Dohlen} \& et
  al.}{{Vigan} et~al.}{2016}]{2016SPIE.9912E..26V}
{Vigan} A.,  {N'Diaye} M.,  {Dohlen} K.,    et al. 2016, {Stop-less Lyot
  coronagraph for exoplanet characterization: first on-sky validation in
  VLT/SPHERE}.
p. 991226

\bibitem[\protect\citeauthoryear{{Wood}, {Boyajian}, {von Braun} \& et
  al.}{{Wood} et~al.}{2019}]{2019ApJ...873...83W}
{Wood} C.~M.,  {Boyajian} T.,  {von Braun} K.,    et al. 2019, \apj, 873, 83

\bibitem[\protect\citeauthoryear{{Zurlo}, {Vigan}, {Galicher} \& et
  al.}{{Zurlo} et~al.}{2016}]{2016A&A...587A..57Z}
{Zurlo} A.,  {Vigan} A.,  {Galicher} R.,    et al. 2016, \aap, 587, A57

\end{thebibliography}

\clearpage
\appendix

\section{Atmospheric retrieval of HD\,19467\,B using the full spectrum}

\begin{figure}
\centering
\includegraphics[width=\columnwidth]{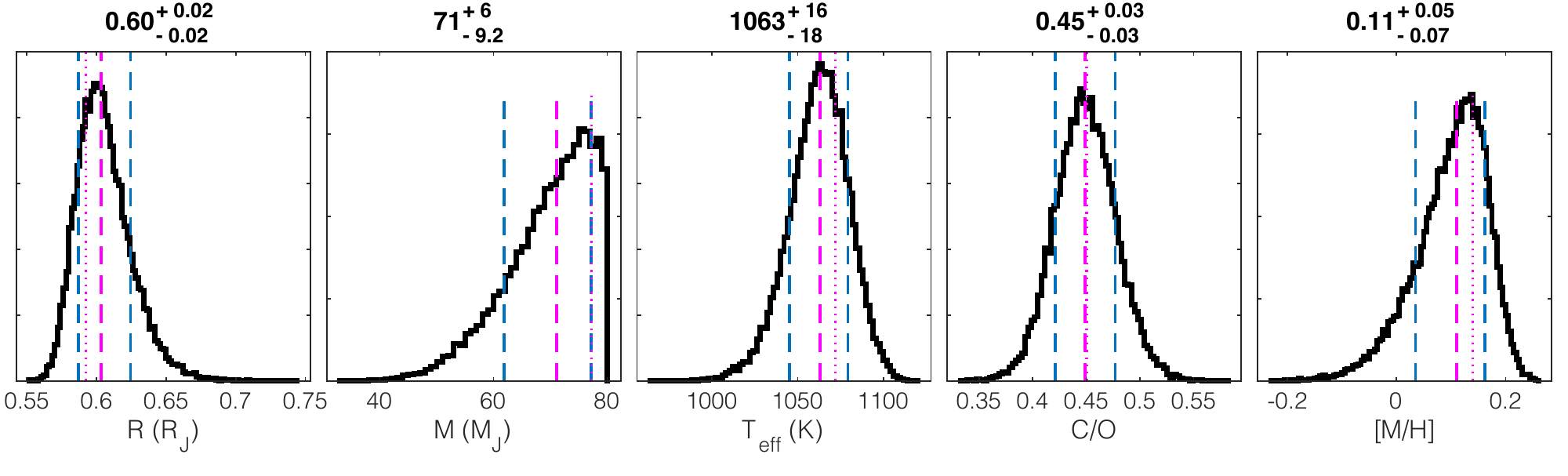}
\includegraphics[width=\columnwidth]{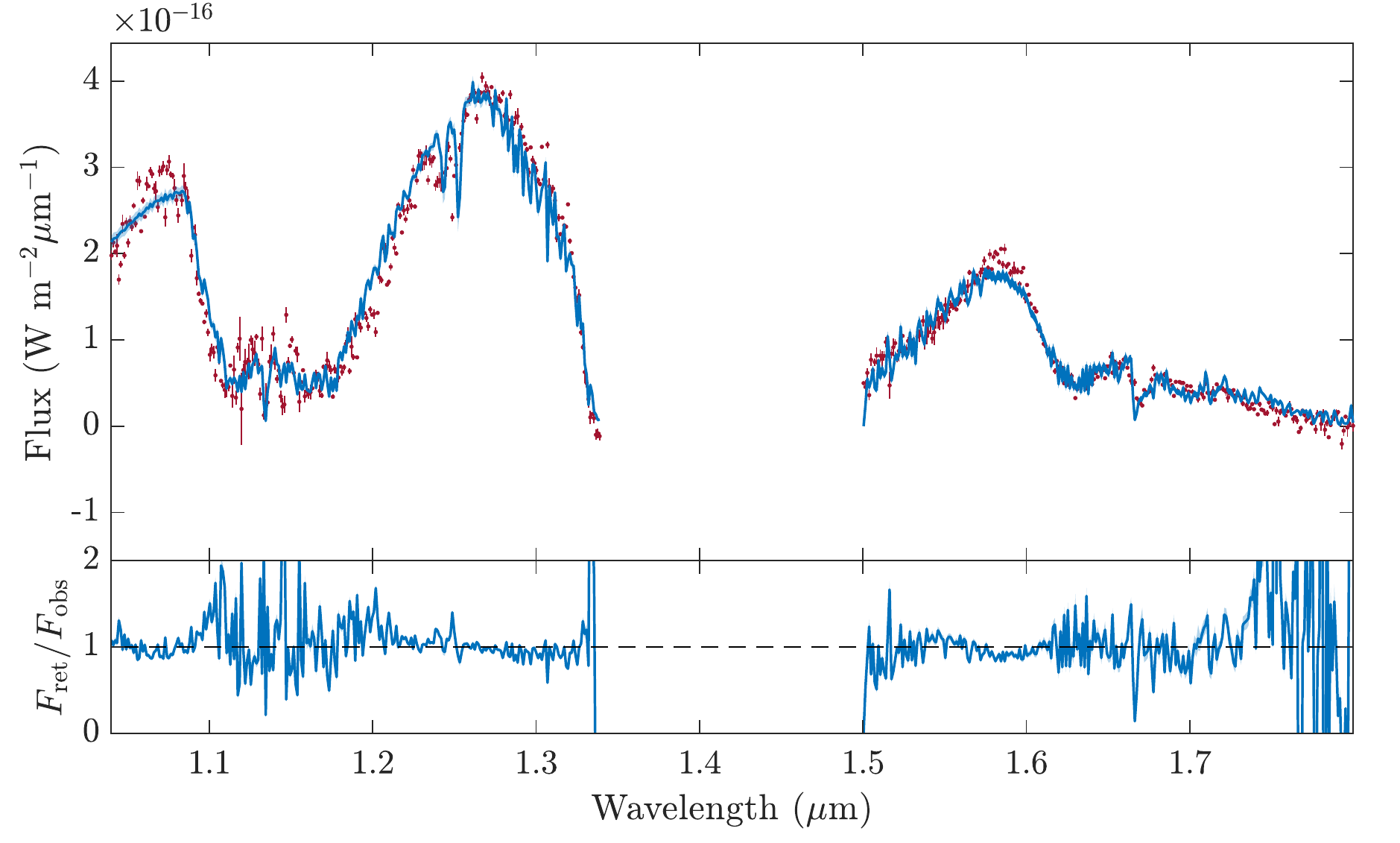}
\caption{{\it Top}: Posterior distributions of derived quantities for the retrieval of HD\,19467\,B. The posteriors are shown for the brown dwarf radius $R$, its mass $M$, effective temperature $T_\mathrm{eff}$, the C/O ratio, and the overall metallicity [M/H]. {\it Bottom}: Posterior spectra and residuals. The solid lines refer to the median of all posterior spectra. The shaded areas signify the 3$\sigma$ confidence intervals of the spectra. The measured spectrum is indicated by the red data points.}
\label{f:HD19467post_add_app}
\end{figure}

\begin{figure*}
\includegraphics[width=0.8\textwidth]{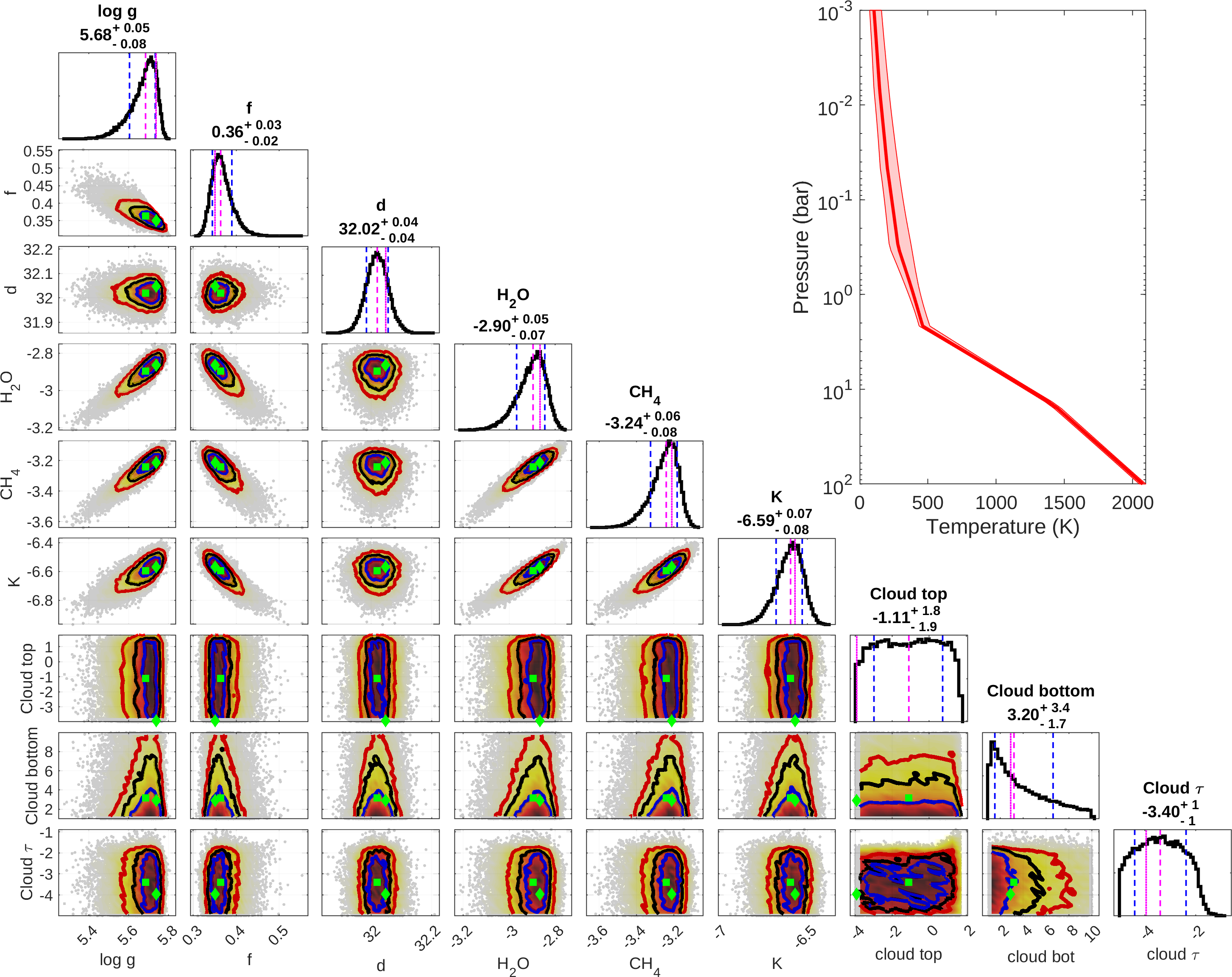}
\caption{Posterior distributions for the retrieval of HD\,19467\,B using the full available spectrum. See caption of Figure \ref{f:HD19467post} for details on the posterior plots.}
\label{f:HD19467post_app}
\end{figure*}

In this Appendix, we show the retrieval for HD\,19467\,B using the entire spectrum, i.e. without cutting it below 1.2 $\mu$m. The resulting posterior distributions are shown in Figure \ref{f:HD19467post_app}, while derived quantities and the posterior spectra are presented in Figure~\ref{f:HD19467post_add_app}.

The results suggest that without removing the problematic region below 1.2 $\mu$m, we obtain much higher surface gravities and too small calibration factors and, therefore, radii. With a value of only $0.6 R_\mathrm{Jup}$, the derived median radius is probably too small for a brown dwarf. While we still obtain sub-solar C/O ratios, the overall metallicity is now super-solar. The posterior spectra shown in bottom panel of Figure \ref{f:HD19467post_add_app} corroborate our previous approach of removing the peak around 1 $\mu$m from our retrieval analysis. It is clear, that the model has problems fitting this spectral region. 

\label{lastpage}

\end{document}